\documentclass[final]{svjour3}
\smartqed  
\usepackage{graphicx}
\usepackage{cite}
\journalname{Bulletin of Mathematical Biology (2009) 71:1160–-1188}
\begin{document}

\title{Ecological Invasion, Roughened Fronts, and a Competitor's Extreme
Advance: \\ Integrating Stochastic Spatial-Growth Models%
\footnote{The original publication is available at \\
www.springerlink.com/content/8528v8563r7u2742/}}

\titlerunning{Roughening of Invasive Fronts}

\author{\mbox{Lauren O'Malley \and G. Korniss \and Thomas Caraco} }


\institute{ Lauren O'Malley  \and G. Korniss
\at Department of Physics, Applied Physics, and Astronomy,\\
Rensselaer Polytechnic Institute, 110 8th Street, Troy, NY 12180-3590, USA \\
\email{omalll@rpi.edu, korniss@rpi.edu}
\and Thomas Caraco (corresponding author)
\at Department of Biological Sciences, University at Albany, Albany NY 12222, USA \\
\email{caraco@albany.edu} }

\date{}

\maketitle

\begin{abstract}
Both community ecology and conservation biology seek further
understanding of factors governing the advance of an invasive
species.  We model biological invasion as an individual-based,
stochastic process on a two-dimensional landscape.  An ecologically
superior invader and a resident species compete for space
preemptively.  Our general model includes the basic contact process
and a variant of the Eden model as special cases.  We employ the
concept of  a ``roughened'' front to quantify effects of
discreteness and stochasticity on invasion; we emphasize the
probability distribution of the front-runner's relative position.
That is, we analyze the location of the most advanced invader as the
extreme deviation about the front's mean position.  We find that a
class of models with different assumptions about neighborhood
interactions exhibit universal characteristics.  That is, key
features of the invasion dynamics span a class of models,
independently of locally detailed demographic rules. Our results
integrate theories of invasive spatial growth and generate novel
hypotheses linking habitat or landscape size (length of the invading
front) to invasion velocity, and to the relative position of the
most advanced invader.
\end{abstract}


\section{Introduction}
Invasive species threaten biodiversity in many ecological communities
\cite{Ruesink_1995,Rosenzweig_2001}, and impose significant costs on agriculture and public
health \cite{Pimentel_2000}.  Spatial advance is the most
obvious phase of invasion \cite{Shigesada_97,Hastings_2005},
and perhaps the most important \cite{Lockwood_2007}. Indeed, the dynamics of
advancing fronts challenges our capacity to predict how
invasive species, emerging infections, and evolutionary adaptations
spread across landscapes \cite{Andow_1990,OMalley_TPB2006}.
Our study emphasizes certain essential
features, signature characteristics, of advancing fronts that
let us integrate some common models of stochastic spatial growth.

We model preemptive competition between an
invader and a resident species
as a stochastic process in a two-dimensional environment. As
the invading competitor advances, the model exhibits variability of
the invader's location about the front's mean position. Generation
of variability along the front separating invader and resident species is
called interface roughening
\cite{Krug_1990,Barabasi_1995,Halpin_1995}. In addition to invasion
processes in population biology \cite{OMalley_PRE2006}, roughened fronts
emerge commonly in other systems,
including the growth of cancerous tumors \cite{Bru_2003}, task-completion landscapes in
in massively parallel computer networks \cite{Korniss_2000,Korniss_2003}, and
surface growth in materials science \cite{Barabasi_1995}.  Models built on
quite different assumptions can roughen in the same way; features dependent on
roughening will then behave similarly across models.

We recently analyzed the time course of roughening in our
two-species model for competitive invasion \cite{OMalley_PRE2006}.  In this paper
we focus on properties of the invader's extreme advance, the
front-runner's location \cite{Ellner_1998}.  We investigate effects
of stochasticity, discreteness and habitat size on the invasion process
(Habitat, or landscape, size refers to the linear extent of the system transverse
to the direction of invasive advance).  For particular parameter
values and initial conditions, our model for spatial invasion
becomes equivalent to the contact process (CP)
\cite{Harris_1974,Durrett_1994ptrsl,Hinrichsen_2000,Oborny_2005}.
Another parameter set produces a variant of the Eden model
\cite{Eden_1961,Jullien_PRL1985,Jullien_JPA1985,Kawasaki_2006};
both CP and the Eden model have served as archetypes for modelling
invasive spread. We aim to extract features common to a family of
models (hence, characteristics of a universality class).  To do so, we first investigate the scaling
behavior of our model's interface width (variability of the invader's advance about its mean position),
both as a function of time and of habitat size.  We then address the
shape of the probability density of the
front-runner's position.  Our analyses reveal consistent universalities.
Model details, of course,
affect non-universal coefficients (or prefactors), but universal
features lead to predictions spanning an important class of stochastic growth models.
Furthermore, although the front's velocity depends on details of a
model's local birth and death rates, the approach to the front's
asymptotic velocity - again, as a function of time and of habitat
size - is universal, and these dependencies offer a useful tool for
identifying ecologically salient properties of a given invasive front.

\subsection{Advancing fronts: background comment}
Spatial expansion characterizes biological invasion from local to geographic scales
\cite{Minogue_1983,vdBosch_1992,Shigesada_97}.  Selecting a framework for modelling
spatially detailed, invasive growth has
been posed as a choice between reaction-diffusion equations for
continuous densities or stochastic discrete (individual-based)
models \cite{Durrett_1994tpb}.  The former include deterministic
reaction-diffusion systems, which often permit approximation
of the invader's speed as the asymptotic
velocity of a travelling wave \cite{Dwyer_1995,Caraco_2002,Murray_2003}.  Significant
generalizations of the basic deterministic theory
vary schedules of reproduction or the distribution
of dispersal distance \cite{Cantrell_1991,Shigesada_1995,Kot_1996,Neubert_2000,Dwyer_2006}
and treat advancing fronts as a travelling waves with
constant or increasing velocity.  However, travelling waves invoke
infinitesimal population densities
\cite{vBaalen_1998}.  The front can be ``pulled''
by growth and spread of the invader at locations where its density
is near 0 \cite{Snyder_2003}; the model may consequently overlook
realistic features of the dynamics of rarity \cite{Lewis_2000,Clark_2003,Antonovics_2006}.
That is, deterministic
reaction-diffusion equations are not
capable of capturing ecologically relevant effects of spatially correlated variability
along the types of fronts we study.

Nevertheless, stochastic partial differential equations (or Langevin
equations) and equivalent field-theoretical descriptions have, for
decades, been employed successfully as coarse-grained (i.e.,
``mesoscopic," continuous-density) representations of both
interacting particle systems in physics and chemistry
\cite{Korniss_1997,Pechenik_1999}, and individual-based models in
biology \cite{Moro_2001,Escudero_2004}. In principle, for every
individual-based model, one could, in the continuous-time limit,
construct a dynamically faithful coarse-grained representation
using a stochastic partial differential equation
\cite{vKampen_1981,Gardiner_1985,Schmittmann_1995,Hinrichsen_2000}.
Deriving the appropriate noise terms presents a challenge, and the
resulting model, with some exceptions, will not likely yield an
analytically tractable solution. One then often resorts, as in the
case of individual-based models, to numerical investigation. The
important point, for questions we address, is that properly
developed Langevin equations
\cite{vKampen_1976,Doi_1976,Peliti_1985} and spatially explicit,
individual-based models should exhibit identical scaling behavior
when critical, large-scale levels of variability govern properties
of the invasion dynamics.  We therefore choose the more familiar
individual-based approach \cite{DeAngelis_1992}.

Individual-based constructions capture
effects of nonlinearity and stochasticity governing invader
dynamics at introduction and at an advancing front \cite{Wilson_1993,Wilson_1998,Thomson_2003}.
For both individual-based and stochastic
partial differential equation models, if
diffusion is limited, the front is usually ``pushed''
at a velocity less than that of the corresponding deterministic
diffusion model \cite{Moro_2003}.  Furthermore, fronts in these
stochastic models roughen; that is, they fluctuate
strongly \cite{Kardar_1986}. Given an initially flat interface
between species, the invader quickly advances different
distances along the length of the front. As the front roughens, the size of random
fluctuations along the front become spatially correlated
\cite{Racz_1988,bAvraham_1998,Majumdar_2004}, and this correlated variation
exerts significant, time-dependent effects on the front's
velocity \cite{Krug_1990,vSaarloos_2003}. Roughening
also implies that the time-dependent position of the invader's
furthest advance, the ``front-runner,'' will
be influenced by the same spatial correlations. One-dimensional models cannot, of course,
generate a roughened front.

Our analysis of the front runner's position
involves the extreme value among
strongly correlated random variables; Fisher and Tippett (1928) and
Gumbel (1958) first derived the distribution of the extreme value
among independent random variables. For roughened interfaces, results
for independent (or weakly-correlated) random variables break down because of the
strongly correlated levels of variability along the front.
Recently Majumdar and Comtet (2004, 2005) analytically obtained
the scaling behavior of the distribution of the extreme advance
of the Kardar-Parisi-Zhang interface \cite{Kardar_1986}.
The interface-set that emerges in
invasion models studied here belongs precisely to this
universality class; hence one immediately obtains the shape of the
probability distribution of the extreme advance (the front-runner) in the steady state.

\section{Model}
Invaders often must overcome competition \cite{Elton_1958,Parker_1998,Hoopes_2002,Simberloff_2002}
to advance spatially.
We assume preemptive competition; neither species can colonize an occupied site until the
occupant's mortality opens that site \cite{Comins_1985,Connonlly_2003,Tainaka_2004,Yurkonis_2004}.
Since we address frontal
advance, we ignore introduction from beyond the habitat; a species
may occupy new sites only through local propagation when one or more
of its nearest-neighboring sites is empty.  Although some invasions may be driven  by
long-distance dispersal \cite{Ferrandino_1996,Lewis_1997,Clark_2001}, our assumption
of limited dispersal has clear empirical justification.  D'Antonio (1993) reports that 98 \emph{per cent} of an
invasive succulent's (\emph{Carpobrotus edulis}) population growth
over four years resulted from spatially clustered clonal propagation.
Invasive Argentine ants (\emph{Linepithema humile}) advance
spatially by locally budding new from existing colonies \cite{Holway_1998}.
For many invasive plants, and some invading animals,
phalanx-like advance should prove realistic within any single
habitat. Importantly, we note that adding explicit finite-range
diffusion, with a small rate, to local colonization does
not change the universal aspects of the quantities studied here
\cite{Moro_2001}.

\subsection{Model dynamics and simulation}
On an $L_x$$\times$$L_y$ lattice, a site represents the minimal resources necessary to
maintain a single individual.   The local occupation number at site
${\bf x}$ is $n_i({\bf x})=0,1$ with $i= 1,2$, referring to the
resident and invader species, respectively.  During a
single time unit, one Monte Carlo step per site [MCSS], $L_xL_y$
sites are chosen randomly for updating.  An empty site may be
occupied by species $i$ through propagation from a neighboring site
at rate $\alpha_i\eta_i({\bf x})$, where $\alpha_i$ is the
individual-level propagation rate for species $i$, and
$\eta_i({\bf x}) = (1/\delta)\Sigma_{{\bf x}'\epsilon {\rm nn}({\bf x})}n_i({\bf x}')$
is the density of species $i$ in the neighborhood around site ${\bf
x}$; ${\rm nn}({\bf x})$ is the set of nearest neighbors of site
${\bf x}$, and $\delta$ is the number of sites in that neighborhood ($\delta=|nn({\bf x})|$).
Unless noted otherwise, the results presented here are for $\delta=4$,
but in a few cases we set $\delta=8$ (Moore neighborhood) and
$\delta=12$ (Moore neighborhood plus 4 sites) for comparison.
An occupied site opens through mortality of the
individual.  We assume density-independent mortality \cite{Cain_1995} and
assign each species the same mortality rate $\mu$.
Summarizing transition rules for an arbitrary site ${\bf x}$,
we have
\begin{equation}
0\stackrel{\alpha_1\eta_1(\bf{x})}{\longrightarrow}1, \;\;
0\stackrel{\alpha_2\eta_2(\bf{x})}{\longrightarrow}2, \;\;
1\stackrel{\mu}{\longrightarrow}0, \;\;
2\stackrel{\mu}{\longrightarrow}0, \;\;
\label{rates}
\end{equation}
where 0, 1, 2 indicates whether a site is open, resident-occupied,
or invader-occupied, respectively.  Table~\ref{table_params} defines the symbols and notation we use.
\begin{table}[t]
\caption{Definitions of model variables and parameters \label{table_params}}
\begin{tabular}{|c|l|}
  \hline
  Symbols & Definitions \\
  $L_x ,\; L_y ( = L)$ & Lattice size \\
  {\bf x} & Location of lattice site \\
  $n_1({\bf x})$ & Occupation number for residents at site ${\bf x}$; $n_1({\bf x})= 0,1$ \\
  $n_2({\bf x})$ & Occupation number for invaders  at site ${\bf x}$; $n_2({\bf x})= 0,1$ \\
  nn({\bf x}) & Set of nearest neighbors around site \textbf{x} \\
  $\delta$ & size of neighborhood around site ${\bf x}$ ($\delta=|nn({\bf x})|$) \\
  $\eta_i({\bf x})$ & Density of species $i$ on nn(\textbf{x}) \\
  $\alpha_i$ & Individual rate of propagule production, species $i$ \\
  $\mu$ & Mortality rate \\
  $\alpha_c (\mu)$ & Minimal propagation rate for persistence \\
  $h_y (t)$ & Rightmost invader in row $y$ at time $t$ \\
  $\overline{h}(t)$ & Mean of $h_y (t)$ \\
  $h_{max} (t)$ & Rightmost invader at time $t$ \\
  $\Delta_{max}(t)=h_{max}(t)-\overline{h}(t)$ & Distance from front-runner to mean of front \\
  $v^*$ & Asymptotic velocity of invasive front \\
  $\langle w^2 \rangle$ & Mean squared interface width \\
  $\xi (t)$ & Correlation length along front \\
  $\alpha$ & Roughness exponent \\
  $\beta$ & Growth exponent \\
  $z$ & Dynamic exponent \\
  ${\rho_i}^*$ & Equilibrium single-species density \\
  $N^*$ & Number of invaders in a row through the width $w$ \\
  \hline
\end{tabular}
\end{table}


We assume that interspecific competition drives the dynamics (i.e.,
each species persists absent competition), and that the invader has
a reproductive advantage.  Therefore, we restrict attention to the
$\alpha_c(\mu)<\alpha_1<\alpha_2$ regime, where $\alpha_c(\mu)$ is
the critical propagation rate below which either species, in the
other's absence, grows too slowly to avoid extinction \cite{Oborny_2005,OMalley_TPB2006}.

\begin{figure}[t]
  \centering
  \vspace*{2.00truecm}
  \includegraphics{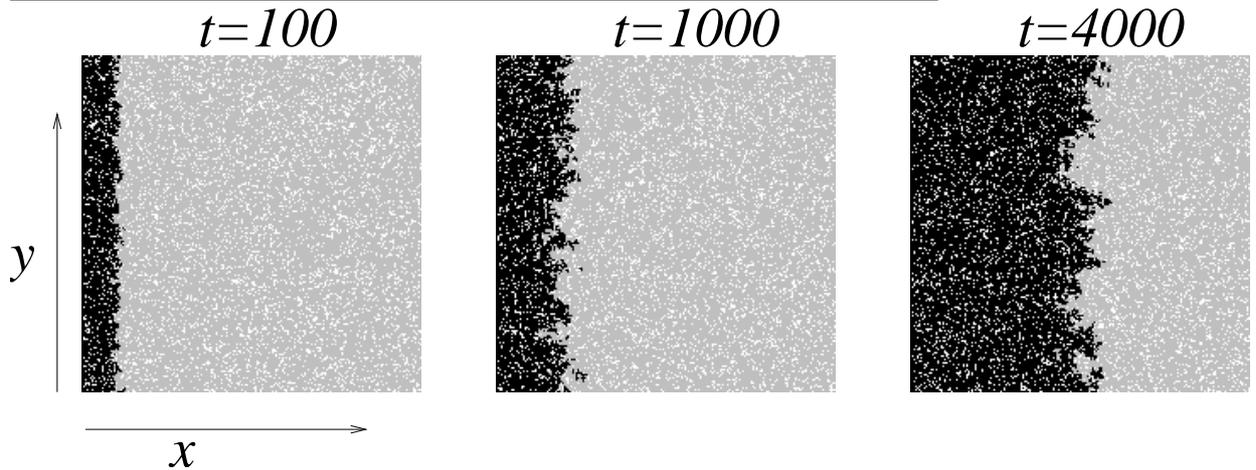}
  \vspace*{4.00truecm}
\caption{\small
Snapshots of advancing fronts for transverse system
size $L$$\equiv$$L_y$$=$$200$ and  $\alpha_1$$=$$0.70$,
$\alpha_2$$=$$0.80$, $\mu$$=$$0.10$. Time has units of
Monte Carlo steps per site (MCSS). White cells represent empty
sites, while gray and black correspond to sites occupied by
residents and invaders, respectively. }
\label{fig1}
\end{figure}

We impose periodic boundary conditions along the $y$-direction of
the $L_x$$\times$$L_y$ lattice.  The initial condition is a flat
linear front (straight vertical line), i.e., the invader completely
occupies a few vertical columns at the left edge of the lattice, and
all remaining sites are occupied by the resident.  Invasive advance,
therefore, proceeds in the $x$-direction.  As a simulation begins,
mortality quickly reduces the density on each side of the front to
the respective ``quasi-equilibrium'' value, where a species'
propagation balances its mortality. As a simulation continues, we
track the location of the invading front by defining the edge as the
location of the right-most individual of the invading species,
$h_y(t)$, for each row $y$ (Fig.~\ref{fig1}). We shall also refer to
this quantity as the local ``height" [borrowing terminology from
the non-equilibrium surface and interface-growth literature
\cite{Barabasi_1995}]. We record the average position
$\overline{h}(t)=(1/L_y)\sum_{y}h_y(t)$ for each time step.  We
estimate velocity once $\overline{h}(t)$ approaches linear increase
with time. We ran each simulation until the front reached the end of the system.
Longitudinal system size $L_x$ has no particular impact on system
behavior.  But the transverse system size $L_y$ plays a fundamental
role in controlling the large-scale properties of the emerging
fronts and the interface region.  Hence the size of the habitat
(specifically, the length of the front) will exert an important
influence on the dynamics of ecological invasion.

If the resident is absent initially, model assumptions
preclude its introduction.  Eliminating the resident species (i.e., $n_1({\bf x})\equiv 0$ for
all ${\bf x}$) reduces the model to the basic contact process for
invaders.  Furthermore, for the specific choice of
$\mu=0$ (i.e., no mortality), the CP itself reduces to a variant of
the Eden model \cite{Jullien_PRL1985}. To emphasize
the strength of the universal aspects of these stochastic growth
processes, we also present some results on the front's extreme advance for these special parameter
choices.

\begin{figure}[t]
  \centering
  \vspace*{3.00truecm}
  \includegraphics{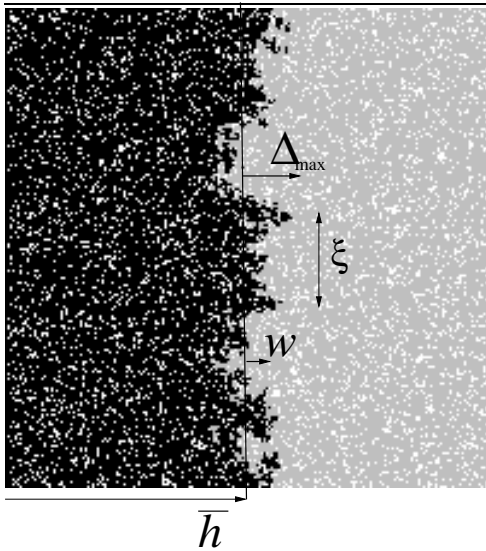}
  \vspace*{4.00truecm}
\caption{\small
Width ($w$) and the extreme advance
($\Delta_{max}$) relative to the mean front position
($\overline{h}$) in a rough front.
For illustration, correlation length $\xi$ is also indicated.}
\label{fig2}
\end{figure}

An advancing front ``roughens'' (Fig.~\ref{fig1}).
That is, the typical size of deviations about the mean front
position increases until the distribution of the mean squared
fluctuation equilibrates statistically.  The resulting steady-state
depends on habitat size $L_y$ \cite{Barabasi_1995,Halpin_1995}.  Figure~\ref{fig2} shows a typical
configuration with ``width'' $w$ about the front's average position $\overline{h}(t)$.
Similar ``wavy" front configurations can be
observed in results plotted in recent front-velocity studies of the basic
CP \cite{Ellner_1998} and of the Eden model \cite{Kawasaki_2006}.
This ``waviness" is at the heart of kinetic roughening, the
focus of our investigation. We emphasize that as habitat size $L_y$
increases, the amplitude of the interface
fluctuations (deviations about the mean) in the steady state increases monotonically.
In turn, one can employ universality arguments and recent results to find the shape
of the distribution of the front-runner's relative position
\cite{Majumdar_2005} and important habitat-size dependent corrections
to the velocity of invasive advance
\cite{Krug_1990}.

\section{Roughening of the front}
Next we apply a framework for dynamic scaling of roughened
interfaces to advance a new perspective on invasion ecology.
To analyze roughening we define the ``width" of the invading front via:
\begin{equation}
w^2(L_y,t) = \frac{1}{L_y}\sum_{y = 1}^{L_y}
[h_y(t) -\overline{h}(t)]^2 \;,
\label{w2}
\end{equation}
Of course, $w^2(L_y,t)$ is itself a fluctuating quantity, so in simulations
we estimate an ensemble average over different runs, or take an average
over a time series in steady state, to obtain $\langle w^2(L_y,t)\rangle$.
Hereafter, we let
$L\equiv L_y$, for convenience, and write $\langle w^2(L,t)\rangle$.
Without creating confusion,
we shall use the notation $w=\sqrt{\langle w^2(L,t)\rangle}$, to refer to the typical size of the interface region.

$w^2(L,t)$ in Eq.~(\ref{w2}) is a function of the local stochastic
invasion heights $h_{y}(t)$,and the $h_{y}(t)$ are not
independent random variables. Starting with a flat interface, a single correlation
length $\xi (t)$ develops along the front (Fig.~\ref{fig2}). This
correlation length grows as a power-law function of time $\xi
(t)\sim t^{1/z}$ \cite{Barabasi_1995}, but once $\xi (t)$ spans the
length $L$ of the system, at a crossover time on the order of
$t_\times\sim L^{z}$, roughening has reached its stationary state.
The heights of invasive advance along the front have become and
remain dependent random variables, complicating analysis
of the invader's maximal advance \cite{Raychaudhuri_2001,Majumdar_2004}. We
let $\langle w^2(L,\infty)\rangle$ represent the mean squared
deviation when its distribution has equilibrated; below we focus
our front-runner analysis on these steady-state fluctuations.

Advancing fronts with different locally structured
propagation and mortality processes may still exhibit the same
dependence of roughening on time, and the equilibrium
variability along the front may exhibit the same dependence on habitat size.  Such
fronts belong to the same ``universality class"; universality
offers powerful simplification and generalization \cite{Cardy_1996,Ferreira_2006}.
O'Malley et al. (2006b) found that our model's roughening
behavior belongs to the KPZ universality class (for
Kardar-Parisi-Zhang; see Kardar et al. 1986).  We summarize those
results, since quantifying roughening allows us to
correct the asymptotic velocity of invasion, and to write
the probability density of the front runner's relative position. Appendix B
provides some further details about the KPZ equation,
a continuous stochastic differential equation, that captures
interface dynamics for a large class of discrete,
stochastic individual-based systems.

\begin{figure}[t]
\centering
\vspace*{3.00truecm}
       \includegraphics{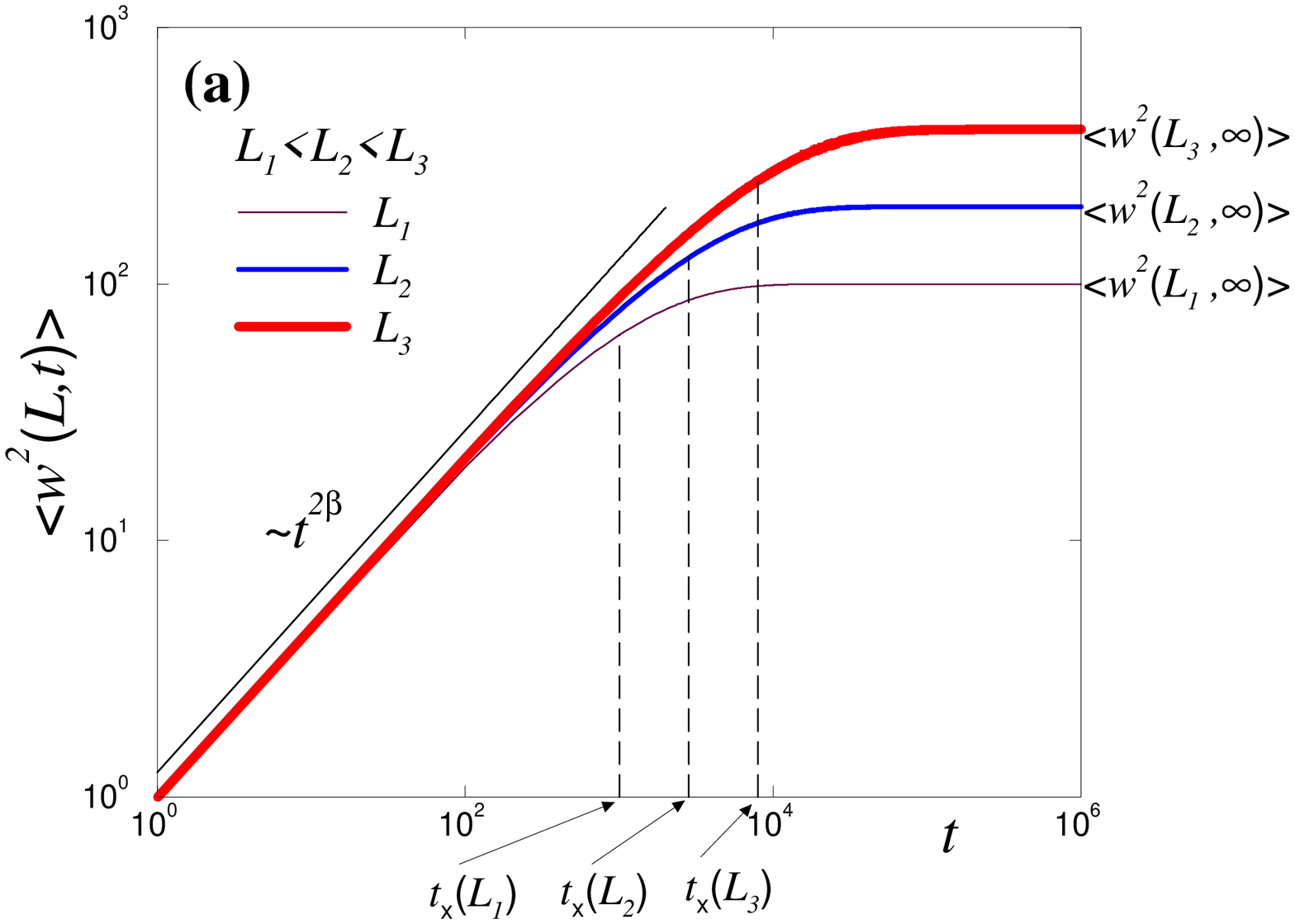}
\includegraphics{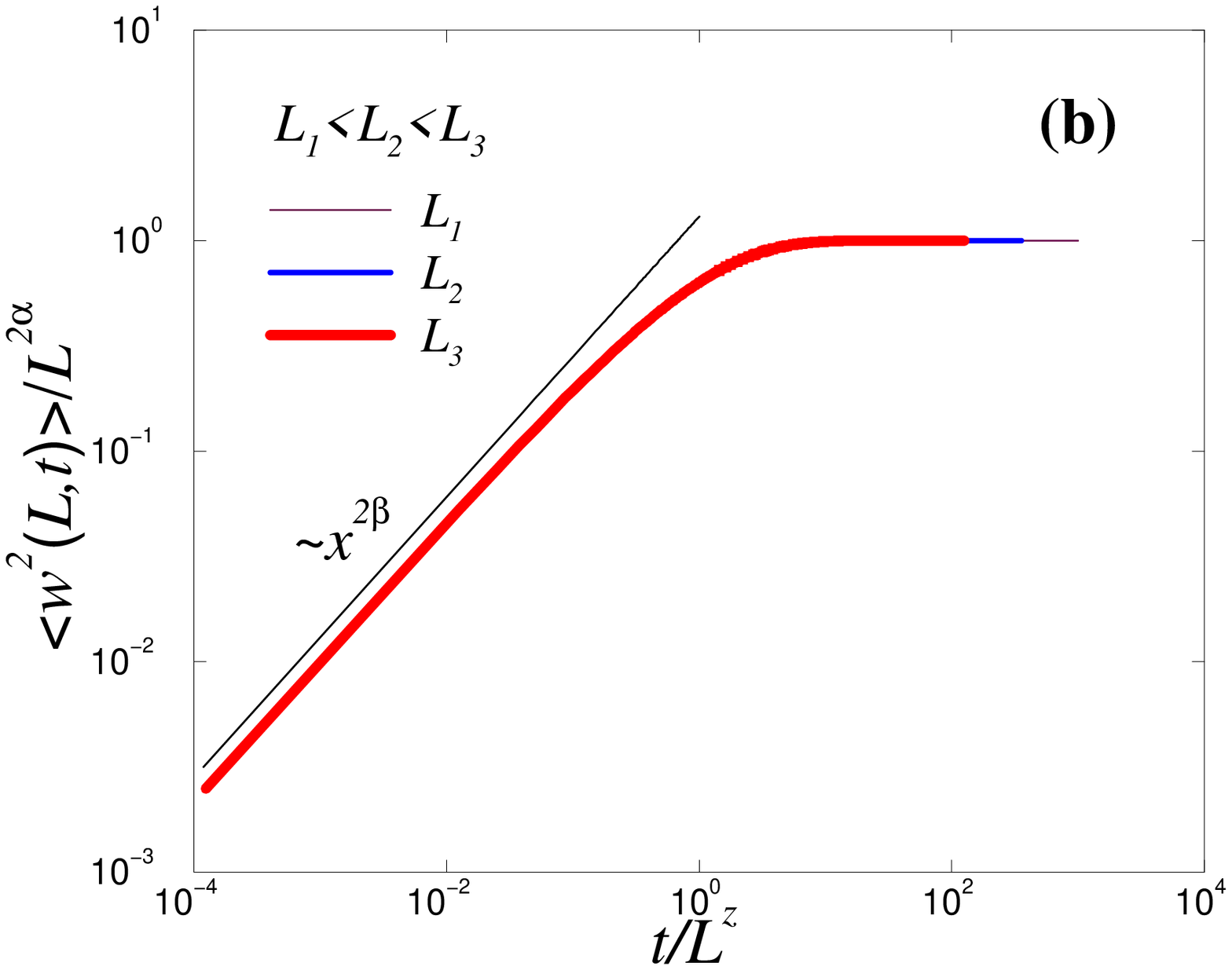}
\vspace*{4.5truecm}
\caption{\small
Schematic plot illustrating the scaling features of the width of rough fronts, Eq.~(\ref{w2_scale}).
(a) Interface width vs time on log-log scales, for different system sizes $L_1<L_2<L_3$.
For early times, the width of the interface increases as
$\langle w^2(L,t)\rangle$$\sim$$t^{2\beta}$.
After a system-size dependent cross-over time $t_\times(L)$$\sim$$L^z$,
the interface fluctuations become stationary, and the width reaches its system-size dependent
steady-state value
$\langle w^2(L,\infty)\rangle$$\sim$$L^{2\alpha}$.
The straight solid line corresponds to the power-law increase for early times before the crossover (growth phase).
For larger transverse system size $L$, it takes longer to reach steady state,
and the width in the steady state is larger.
(b) Scaled plot $\langle w^2(L,t)\rangle/L^{2\alpha}$ vs $t/L^{z}$, yielding full data collapse for different system sizes,
the signature of dynamic scaling \cite{Barabasi_1995}. The straight solid line indicates the power-law behavior $x^{2\beta}$
of the scaled variable $x$$=$$t/L^{z}$ for $x\ll1$.}
\label{roughening}
\end{figure}

The profound nature of roughened fronts (or self-affine interfaces)
lies in the intimate connection between their temporal and spatial
scaling behavior \cite{Barabasi_1995,Halpin_1995}. For a system
with transverse system size $L$, for early times, the width
(or spread) of the interface [Figs.~\ref{fig1},\ref{fig2}] increases in a power-law fashion
$\langle w^2(L,t)\rangle$$\sim$$t^{2\beta}$.
Then, after a system-size dependent cross-over time
$t_\times$$\sim$$L^z$, the front reaches its steady-state: interface
fluctuations about the steadily advancing mean front become
stationary, but their magnitude will depend on the transverse system
size. In particular, the interface width ``saturates" in time, and
its stationary value will scale with the system size as
$\langle w^2(L,\infty)\rangle$$\sim$$L^{2\alpha}$.
Thus, both the time to reach steady state and the size of the
steady-state width depends on the transverse system size $L$ in a
power-law fashion. Put simply, if the transverse system size is
increased, both the time to reach steady state and the width of the
interfacial region in the steady state increase. The exponents
$\alpha$$>$$0$, $\beta$$>$$0$, and $z$$>$$0$ are referred to as the
roughness, growth, and dynamic exponent, respectively. (The
roughness exponent $\alpha$ should not to be confused with the local
propagation rates of the two species.) These fundamental scaling
properties of roughening can be summarized as
\begin{equation}
\langle w^2(L,t)\rangle \sim \left\{
\begin{array}{ll}
t^{2\beta}       & \mbox{for $t \ll t_{\times}$} \\
L^{2\alpha}      & \mbox{for $t \gg t_{\times}$}
\end{array} \right. \;,\;\;\;\;\;\;\;\;\;\;\; t_\times \sim L^z
\label{w2_scale}
\end{equation}
and are illustrated schematically in Fig.~\ref{roughening}. In
general, these exponents obey a scaling relationship
$\alpha=\beta z$ \cite{Barabasi_1995}.

Models whose kinetic roughening can be described by the same
set of exponents are said to belong to the same {\em universality
class} \cite{Barabasi_1995,Halpin_1995}. Fronts generated by models with
different local rules often belong to the same universality class
(for example, fronts in the Eden model, the contact process, and the
two-species model of this paper).  Importantly, models formulated to
address completely different questions, ranging from surface physics and
computer science \cite{Korniss_2003} to ecology, can belong to the
same universality class.

\begin{figure}[t]
\centering
\vspace*{2.50truecm}
       \includegraphics{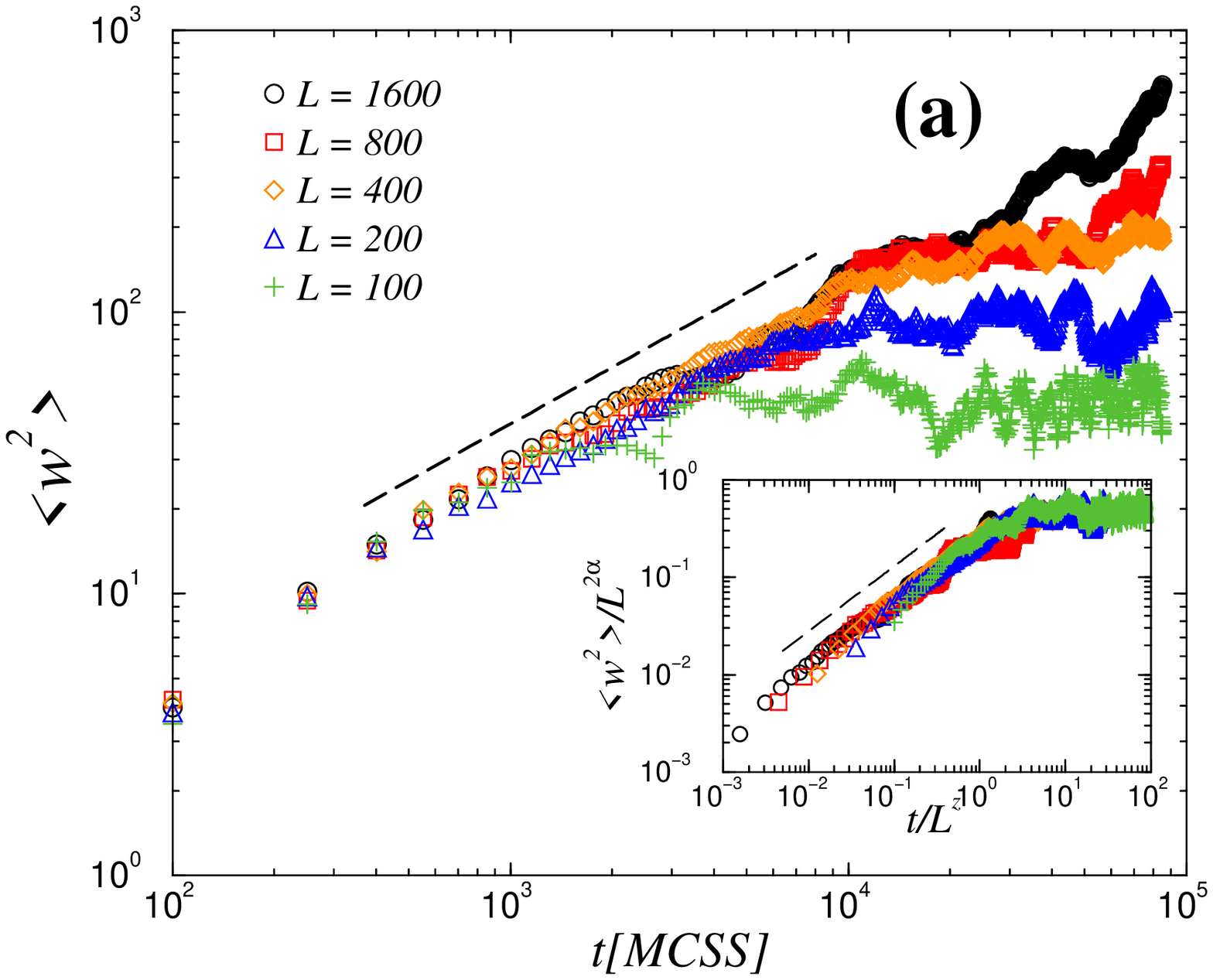}
       \includegraphics{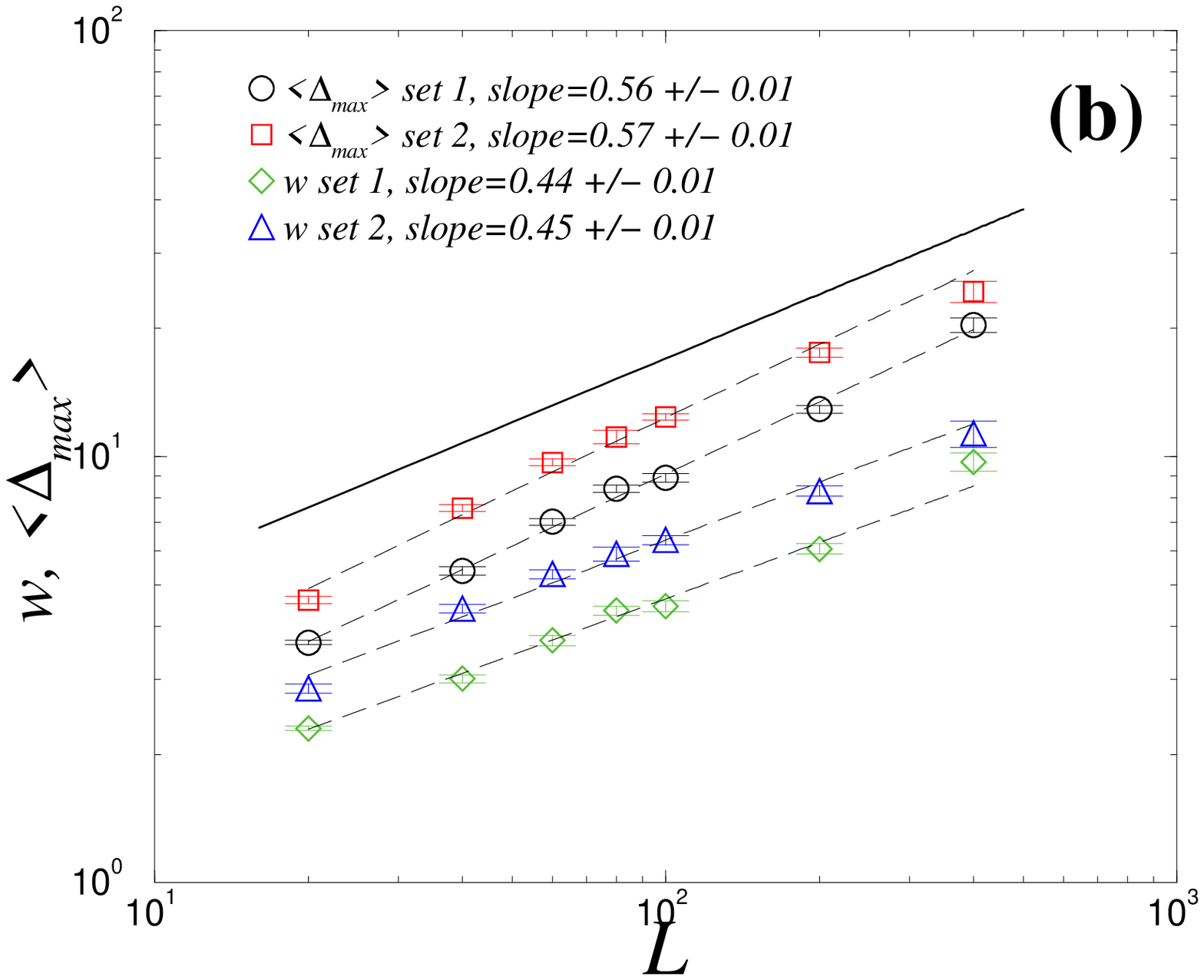}
\vspace*{3.2truecm}
\caption{\small
       (a) Average width as a function of time
       (log-log scales) for various
       system sizes, averaged over $20$ independent realizations with
       $\alpha_1$$=$$0.70$, $\alpha_2$$=$$0.80$, and $\mu$$=$$0.10$.
       Dashed line corresponds to the one-dimensional KPZ power law
       with the exponent $2\beta=2/3$. Inset shows the scaled plot,
       $\langle w^2(L,t)\rangle/L^{2\alpha}$ vs. $t/L^z$
       using the KPZ exponents.
       (b) Steady-state width, $w=\sqrt{\langle w^2(L)\rangle}$,
       and extreme fluctuations, $\langle\Delta_{max}(L)\rangle$,
       (averaged over time series in the steady-state)
       as a function of system size, $L (\equiv L_y)$, for two sets of parameters:
       Set~1 ($\alpha_1$$=$$0.50$, $\alpha_2$$=$$0.70$, and $\mu$$=$$0.20$); and
       Set~2 ($\alpha_1$$=$$0.70$, $\alpha_2$$=$$0.80$, and $\mu$$=$$0.10$).
       Dashed lines indicate the best-fit power laws. The solid line corresponds to the
       the theoretical asymptotic scaling in the KPZ universality class with roughness exponent $\alpha$$=$$1/2$.}
\label{fig3}
\end{figure}

The front in our two-species individual-based model exhibits
scaling consistent with Expression \ref{w2_scale}; the exponents are
$\beta$$=$$1/3$, $\alpha$$=$$1/2$, and
$z$$=$$3/2$ \cite{OMalley_PRE2006}.  Consequently, the front belongs to the so called
KPZ-interface universality class (in one transverse dimension)
(Fig.~\ref{fig3}). Slight deviations from the theoretical exponents
result, in part, from corrections to scaling and ``intrinsic width"
effects for limited system sizes \cite{Moro_2001,Blythe_2001}.
Furthermore, the strong correlation in the steady-state time series of
$w^{2}(t)$ (the auto-correlation time also diverges with the system
size as $L^{z}$) increases sampling error for larger system sizes.
For the KPZ universality class, the analytical result by Foltin et
al. (1994) specifies the universal shape of the distribution of the
steady-state width (see Appendix B.2). The results can be used to
identify the universality class of a model with a rough front; doing
so provides further support that our model belongs to the KPZ class
(Appendix B.2).

\subsection{Roughening and the asymptotic invasion velocity}
Estimating the speed at which an invasive species
advances remains a focus of ecological theory
\cite{Mollison_1995,Ellner_1998,Lewis_2000,Kawasaki_2006}.
Finding an analytical expression for a stochastic front's
asymptotic velocity $v^{*}$
presents a challenge \cite{Pechenik_1999}, particularly
for two-dimensional fronts. Analytically tractable,
pulled fronts emerge naturally in the deterministic
reaction-diffusion limit of our model
\cite{Fisher_1937,Holmes_1994,Dwyer_1995,Murray_2003,vSaarloos_2003} (see Appendix A).  But
their velocity typically overestimates that of the actual
individual-based model in the diffusion-limited regime
\cite{Moro_2001,Moro_2003,Doering_2003}. Our model,
with only local dispersal and no ``explicit" diffusion, lies
precisely in that regime: since interspecific mixing is insufficient about the
interface, and internal fluctuations are important,
deterministic equations break down \cite{Antonovics_2006,OMalley_PRE2006}.

Deterministic reaction-diffusion models neglect discreteness of
individuals, the fundamental source of endogenous
fluctuations \cite{Escudero_2004}.  Consequently,
deterministic reaction-diffusion theory neglects the
substantial, correlated random variation in the extent of the invader's advance along the
 front. To demonstrate the significance
of this difference, we compared results from
Monte Carlo simulation of the individual-based
model defined by the transition rates in Eq.~(\ref{rates}) to
the velocity of travelling waves (pulled fronts) arising in our model's
deterministic reaction-diffusion limit,
$v_{p} = (\mu/\alpha_1) \sqrt{\alpha_2 (\alpha_2 - \alpha_1)}$ (see Appendix A).
Asymptotic invasion velocity $v_{p}$ increases as either
the common mortality rate $\mu$ or competitive asymmetry,
($\alpha_2 - \alpha_1 $), increases; a strong resident competitor
 can slow an invasive species' advance under
a variety of assumed dynamics \cite{Hosono_1998,Gandhi_1999,Weinberger_2002,OMalley_SPRINGER}.
For further comparison with our stochastic model, note that the travelling
wave approaches its steady-state value as $v(t)=v_{p}-{\cal
O}(1/t)$; that is, the travelling wave's temporal correction decays
as $t^{-1}$ \cite{Holmes_1994}.

\begin{figure}[t]
\centering \vspace*{2.50truecm}
       \includegraphics{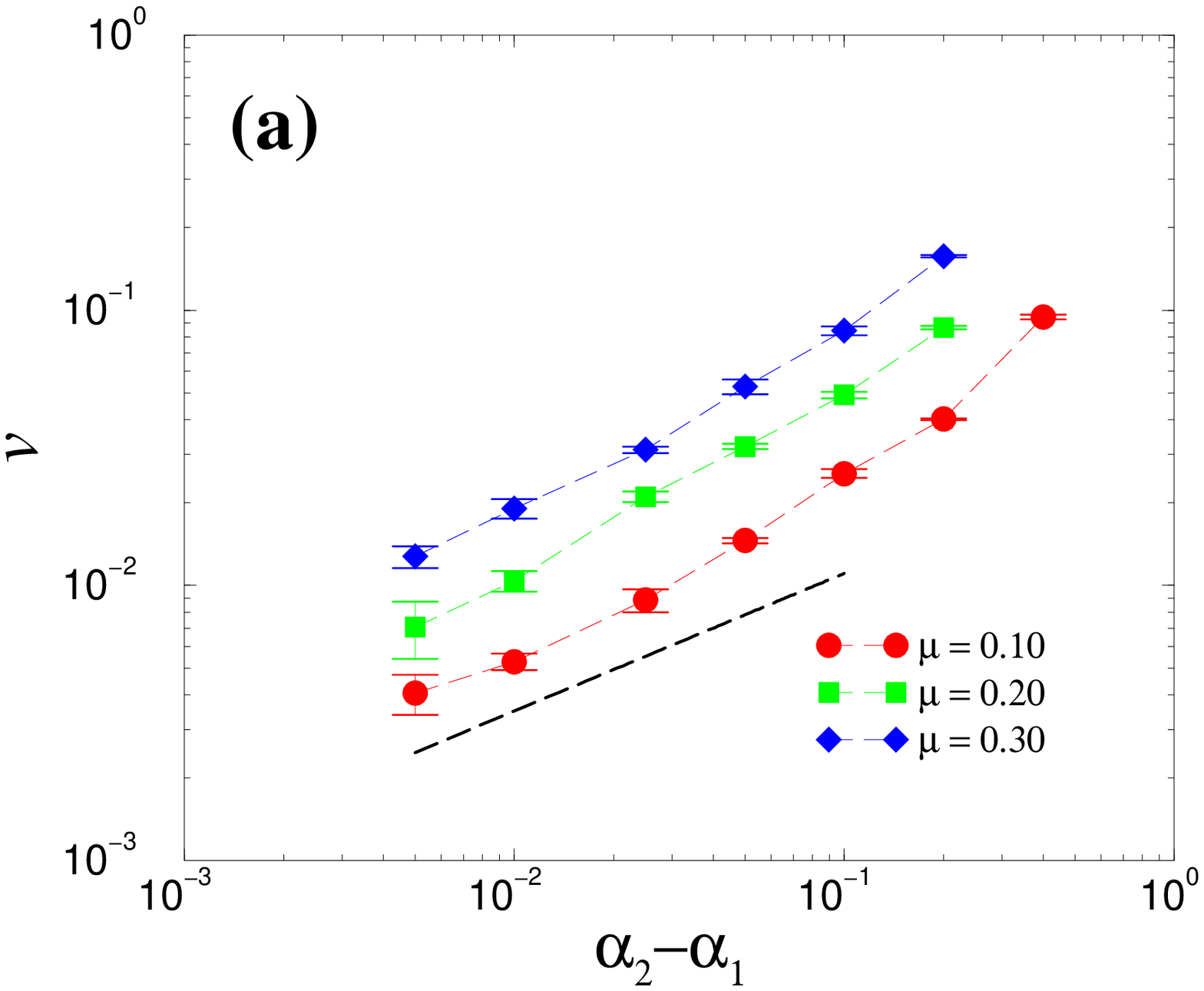}
       \includegraphics{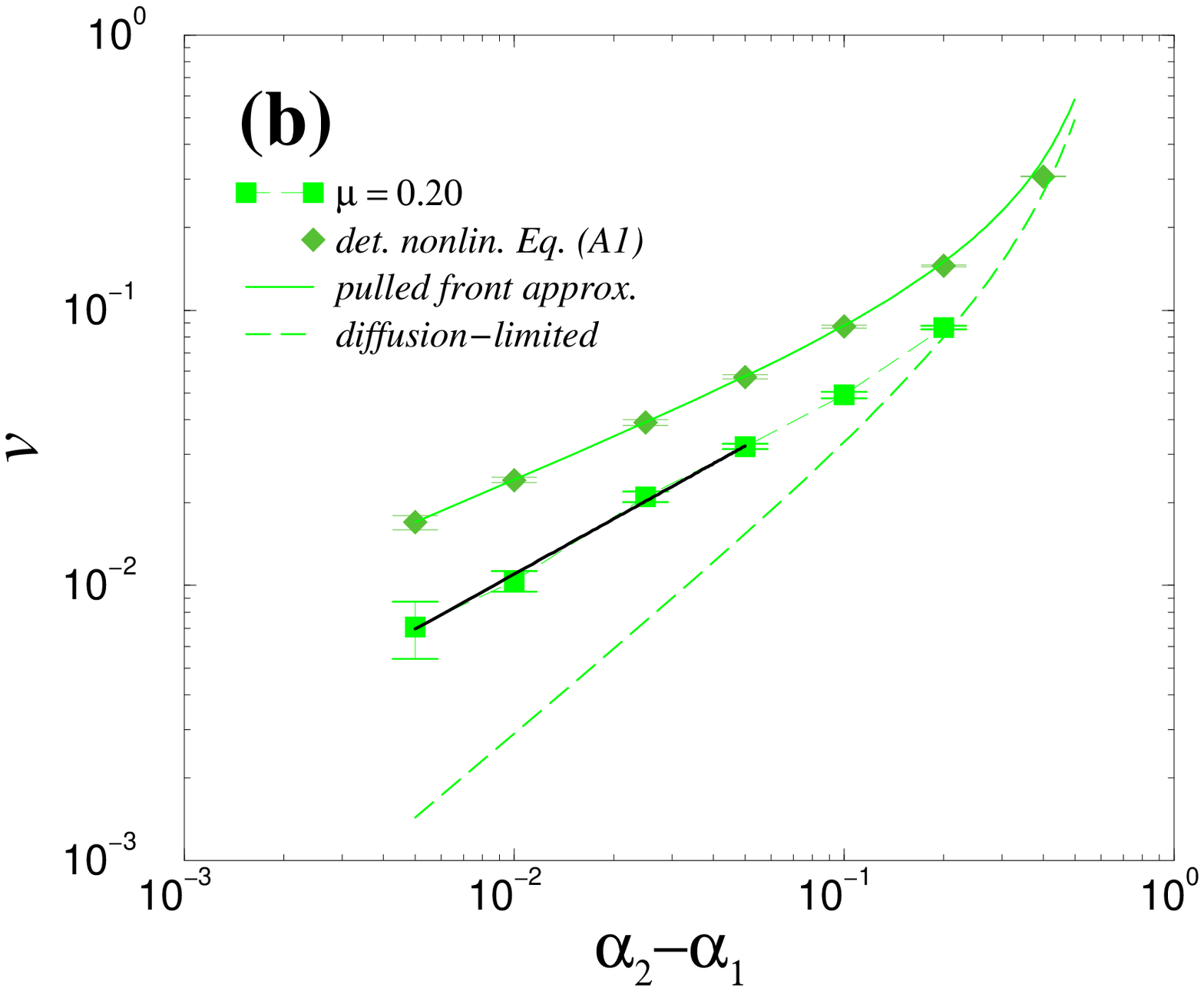}
\vspace*{3.2truecm}
\caption{\small (a) Asymptotic front velocities
from Monte Carlo simulations of the individual-based stochastic
model for fixed $\alpha_2$=$0.70$, as a function of the difference
between propagation rates, $\alpha_2$$-$$\alpha_1$, with
$L_{y}=100$, for three values of $\mu$. (b) Simulation results from
(a) for $\mu=0.20$ compared with the results from numerically iterating the deterministic nonlinear
reaction-diffusion equations, Appendix A, Eq.~(\ref{lattice_RD}) \cite{OMalley_PRE2006,OMalley_SPRINGER}
and with two approximate analytic limits: pulled-front approximation [solid curve,
Eq.~(\ref{velocity_pulled})] and diffusion-limited front [dashed
curve, \cite{Moro_2003}]. The solid straight line is the best-fit
effective power law in the region where the difference between
propagation rates is small, corresponding to $v\sim(\alpha_2
-\alpha_1)^{\theta}$ with $\theta=0.66\pm0.04$. }
\label{fig4}
\end{figure}

Figure~\ref{fig4}(a) plots asymptotic velocities from simulation of
our individual-based model (for fixed $\alpha_2$) as a function of
competitive asymmetry $(\alpha_2 - \alpha_1 )$.  For parameter
values chosen, the invader advances more slowly than our model's
deterministic reaction-diffusion limit (i.e., slower than $v_{p}$)
[Fig.~\ref{fig4}(b)]. It is important to note that the pulled front
approximation $v_p$ can fully reproduce the numerical solutions of
the {\em deterministic nonlinear} reaction-diffusion system
[Appendix A, Eqs.~(\ref{lattice_RD}) and (\ref{continuum_RD})] \cite{OMalley_PRE2006,OMalley_SPRINGER}.
What it fails to account for are the effects of {\em stochasticity}.
In our discrete, stochastic model
the front becomes pushed \cite{vSaarloos_2003}; invasion velocity
depends on the full frontal region, with its nonlinear interactions,
rather than depending on the leading edge only (a pulled front).
For comparison, we also plot an approximate expression for the
velocity, suggested by Moro (2003), $v_{DL}=\mu(\alpha_2/\alpha_1
-1)$, applicable to strongly diffusion-limited systems (see next
section), where the effective diffusion coefficient is negligible
compared to the effective reaction rates. As Fig.~\ref{fig4}(b)
shows, except for large $(\alpha_2-\alpha_1)$, this approximation
underestimates the front velocity of our individual-based model.
This may be expected by noting that the effective reaction and
diffusion coefficients are of the same order of magnitude (see
Appendix A).

The functional expression for a model's asymptotic velocity cannot
be universal, since it must depend on details of the
local dynamics.  However, Krug and Meakin (1990) showed that
temporal and finite-size ($L_y=L$) corrections to the asymptotic velocity are universal within
a given class (see Appendix B.3); a general ecological prediction consequently emerges.
Frontal velocity initially increases
until reaching steady state.  Prior to the steady state, the difference between
the current and asymptotic velocities (the temporal correction) scales similarly with
time for every member of a particular universality class.  More importantly,
the second correction recognizes that steady-state velocity depends on
habitat size; the habitat-size correction is proportional to $L^{-1}$
[see Appendix B.3, Eq.~(\ref{velocity_corr})].  That is,
a given invasive species' front will advance faster as
linear habitat size increases.  More generally, the finite-size correction
to the asymptotic velocity scales similarly with habitat size for
each stochastic model where invasive advance exhibits a KPZ interface.

\subsection{Invader spatial profiles and diffusion limitation}
\begin{figure}[t]
\centering \vspace*{2.50truecm}
       \includegraphics{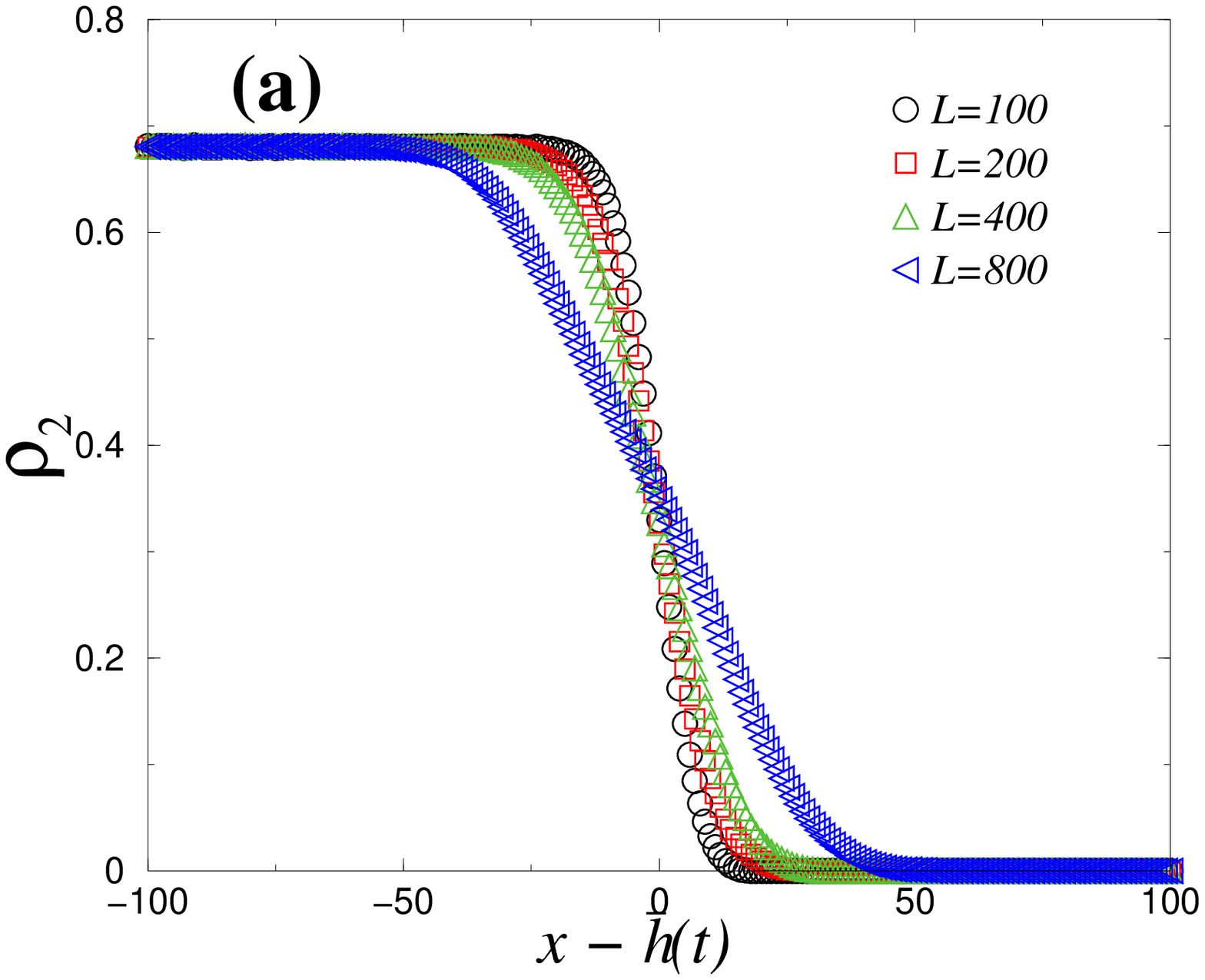}
       \includegraphics{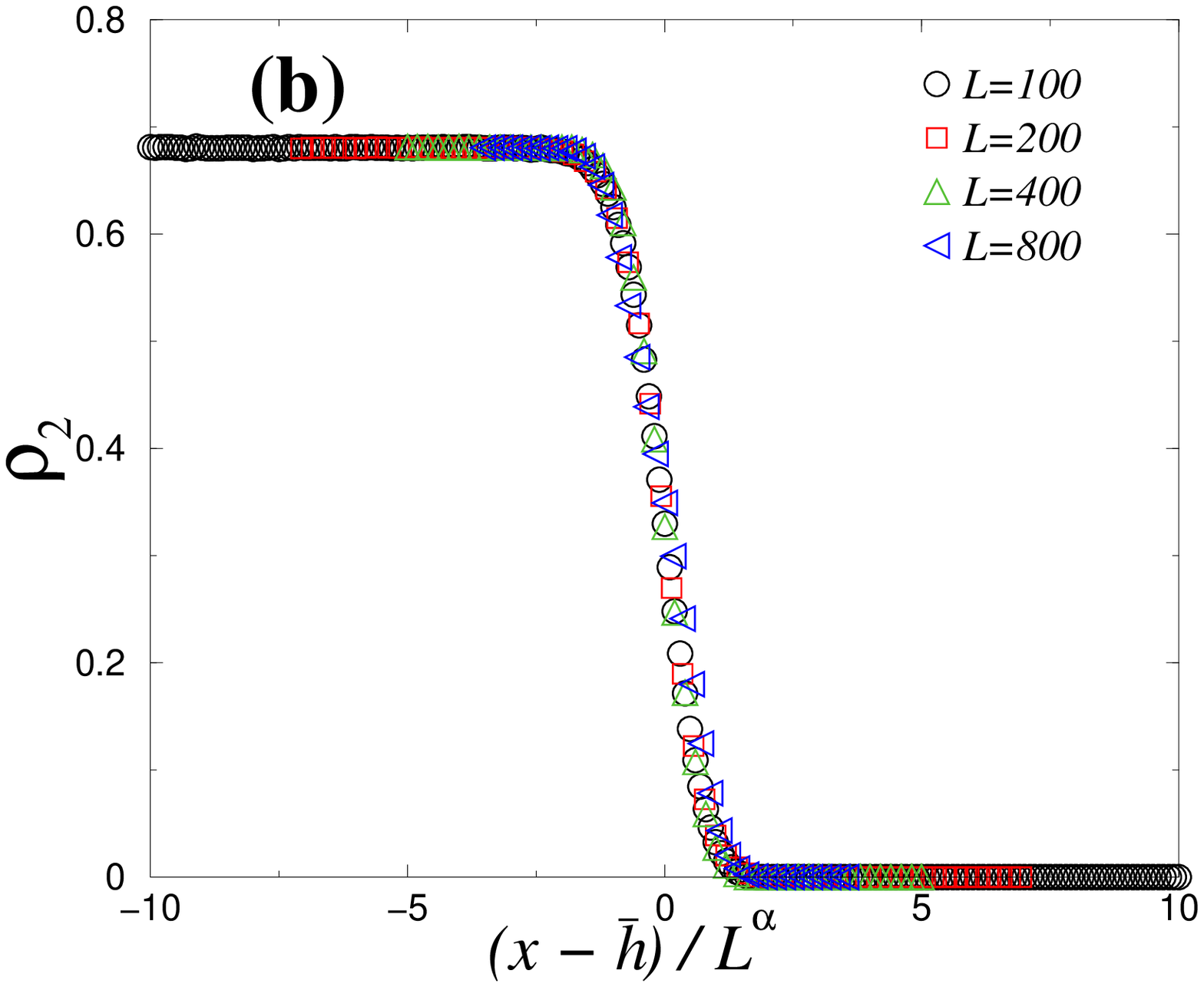}
\vspace*{3.2truecm}
\caption{\small
(a) Steady-state spatial density
profiles of the invader species [averaged over all rows, Eq.~(\ref{density_profile})]
for several habitat sizes, for parameter ``Set 1"
($\alpha_1$$=$$0.50$, $\alpha_2$$=$$0.70$, $\mu$$=$$0.20$), averaged over $10^4$ time
steps. (b) Same data as in (a), but scaled according to the system-size dependent width of
the front ($w\sim L^{\alpha}$ with $\alpha = 1/2$). }
\label{fig7}
\end{figure}

Next we focus on invader density within the width of the
advancing front, to compare stochastic
and deterministic invasion models in a general context. We constructed
 density profiles for the invader along the horizontal
direction (averaged over rows $y$, for a given $x$):
\begin{equation}
\rho_{2}(x) = \frac{1}{L_y}\sum_{y=1}^{L_y}n_{2}(x,y)\;.
\label{density_profile}
\end{equation}
The density profile should spread horizontally in
response to the front's roughening (as a function of time during growth
phase, and as a function of habitat size in steady state). Figure~\ref{fig7}(a)
shows the invader's steady-state density profiles
as a function of distance from the front's mean
position.  Profiles are averaged over $10^4$ time steps, and plotted for
a series of habitat sizes $L$.  Behind the front, where the resident has been excluded,
invader density $\rho_2 (x)$ rests at its single-species
equilibrium ${\rho_2}^*$. Invader density declines through the width,
and only the resident occupies locations well right of the front.

The steady-state density profiles show that the span of the frontal region,
within which ${\rho_2}^* > \rho_2 (x) > 0$, increases progressively with
habitat size $L$ [Fig.~\ref{fig7}(a)].  Since steady-state roughening
scales with habitat size according to $\langle w^2 (L, \alpha
)\rangle \sim L^{2 \alpha}$, rescaling the abscissa by $L^\alpha$ ($ = L^{1/2})$
results in the different density profiles falling on the same plot [Fig.~\ref{fig7}(b)],
an articulate data collapse indicating the fundamental, underlying dependence
(cf. Antonovics et al. 2006).

We stress that expansion of density
profiles with increasing habitat size is not a result of greater
mixing of the two species
within the span of the front.  Rather, density profiles result from
emerging spatial correlations between fluctuations in the invader's advance along the length of the front,
a collective result of the neighborhood-level stochastic dynamics.
Indeed, spatial configurations in Figs.~\ref{fig1}
and \ref{fig2} reveal that few single rows, if any, suggest an
invader-density gradient.  In fact, the density profile in most rows
resembles a ``shock-wave".

Moro (2003) analyzes transitions
from pushed-front to pulled-front velocities in discrete, stochastic
models.  Our simulations fall into Moro's diffusion-limited regime,
where strong spatial correlations produced by local
propagation dominate effects of mixing along the front.  Our model's
assumptions (in particular, no explicit diffusion)
and parameter values imply that the mean-field assumption of strong
mixing at the between-species interface cannot hold. Consequently,
our competitive invasion produces, in any given row,
an invader-density gradient that spans only a narrow region; essentially, the
invading species advances as a shockwave front.
Moro's (2003) analysis identifies the significance of
$N^*$, the (approximate) number of invaders occupying a row of the
interface between the two regions of single-species equilibrium densities.
The front's velocity asymptotically approaches the
reaction-diffusion (pulled-front) velocity $v_{p}$ only as $N^*
\rightarrow \infty$.  For a mortality rate of $\mu$ =
0.1, our simulations have 1 $< N^* <$ 2, so that the impact of discreteness and
stochasticity remains strong.

\section{The front-runner: roughening and scaling of extremes}
Finally, we turn to the problem of the invading species' ``front-runner''
\cite{Ellner_1998,Clark_2001,Thomson_2003},
and scaling of the distribution of the maximal invasive incursion.
The front-runner's position
presents an extreme value problem, complicated by the
probabilistic, strongly correlated dependence of the invader's
advance at different locations along the front \cite{Raychaudhuri_2001,Majumdar_2004}.
Hence, traditional extreme-value statistics
\cite{Fisher_1928,Gumbel_1958,Berman_1964,Galambos_1987,Galambos_1994} are not applicable.

We restrict attention to height fluctuations occurring after
roughening has equilibrated ($t>>t_{\times}$$\sim$$L^{z}$).
At each time step we identify the
farthest-advanced location of the invading species,
$h_{max}(t)=\max_{y}\{h_{y}(t)\}$.  Measuring the extreme advance
relative to the front's mean position, $\overline{h}(L,t)$, we
obtain the maximal relative height at time $t$, $\Delta_{max}(t) =
h_{max}(t) - \overline{h}(L,t)$. We sample $\Delta_{max}(t)$
sufficiently to construct histograms to estimate the probability
density of the extreme fluctuations $P(\Delta_{max}, L)$, for
different propagation/mortality rates, and for different habitat
sizes $L$.

Results presented above led us to
conclude that the front in our two-species invasion model belongs
to the KPZ class. Hence, we can proceed to
characterize properties of invasive extremes. Given steady-state
roughening of the invading species' front, the probability density of the extremes
can be written as
\begin{equation}
P(\Delta_{max},L)= \frac{1}{\langle\Delta_{max}\rangle}\Psi(\Delta_{max}/\langle\Delta_{max}\rangle)\;,
\label{P_ext_hist}
\end{equation}
where $\Psi(u)$ is the probability density of the scaled variable
$u=\Delta_{max}/\langle\Delta_{max}\rangle$ and is independent of habitat-size $L$.
$\Psi(u)$ is a scaled Airy density for all interfaces belonging to the KPZ universality
class \cite{Majumdar_2004,Majumdar_2005} (see Appendix~B.4).
Importantly, the steady-state average of the extreme advance scales
with the habitat size in the same fashion as the width $w$, i.e.,
$\langle\Delta_{max}\rangle\sim L^{\alpha}$ with
$\alpha$$\approx$$1/2$ [Fig.~\ref{fig3}(b)]. That is, the expected
displacement of the front-runner ahead of the invader's mean
position increases with habitat size in the above power-law fashion.
Again, certain system-specific characteristics still depend on local rates,
neighborhood size, etc., but our general ecological prediction does not depend
on these details.

\begin{figure}[t]
\centering
\vspace*{2.50truecm}
       \includegraphics{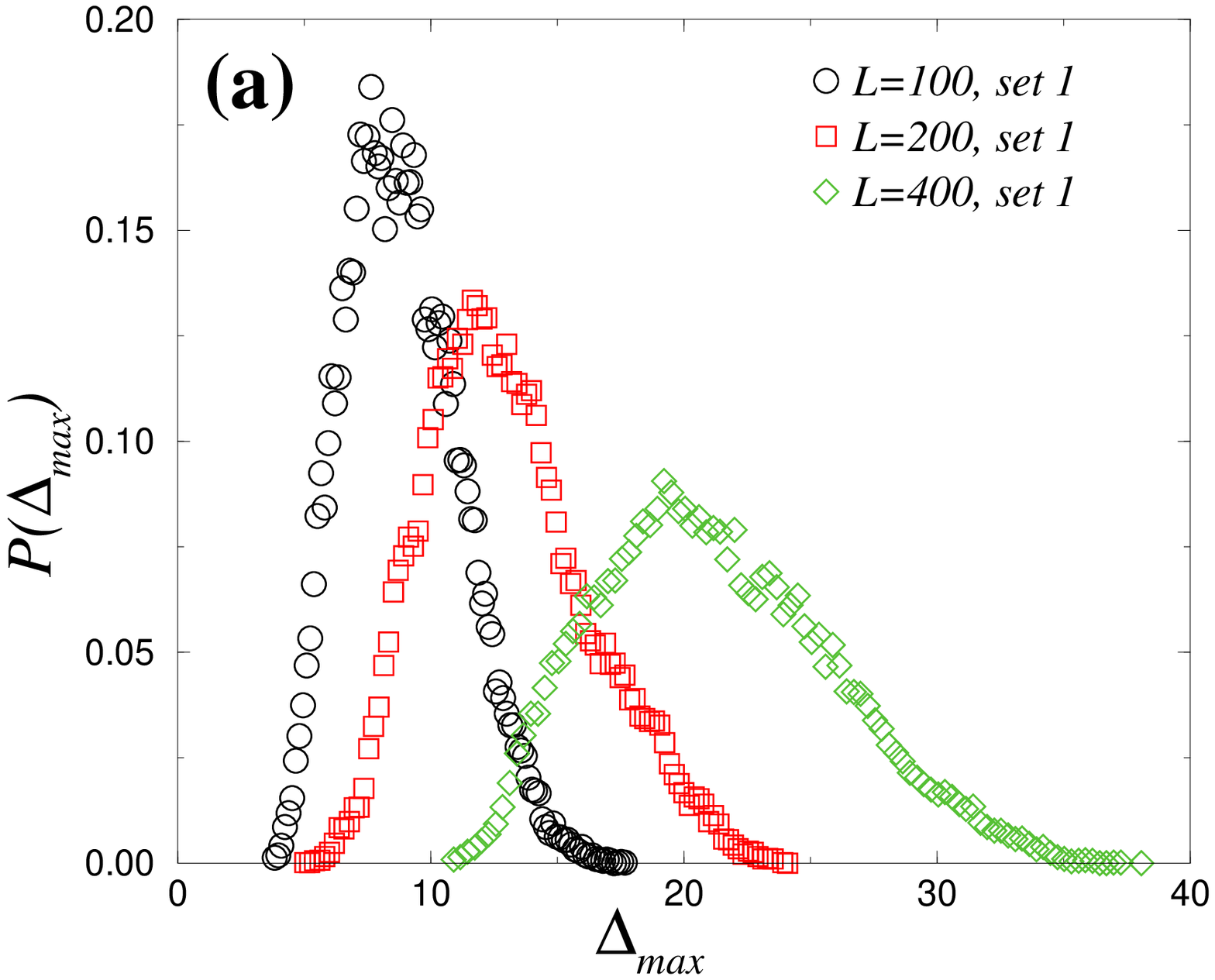}
       \includegraphics{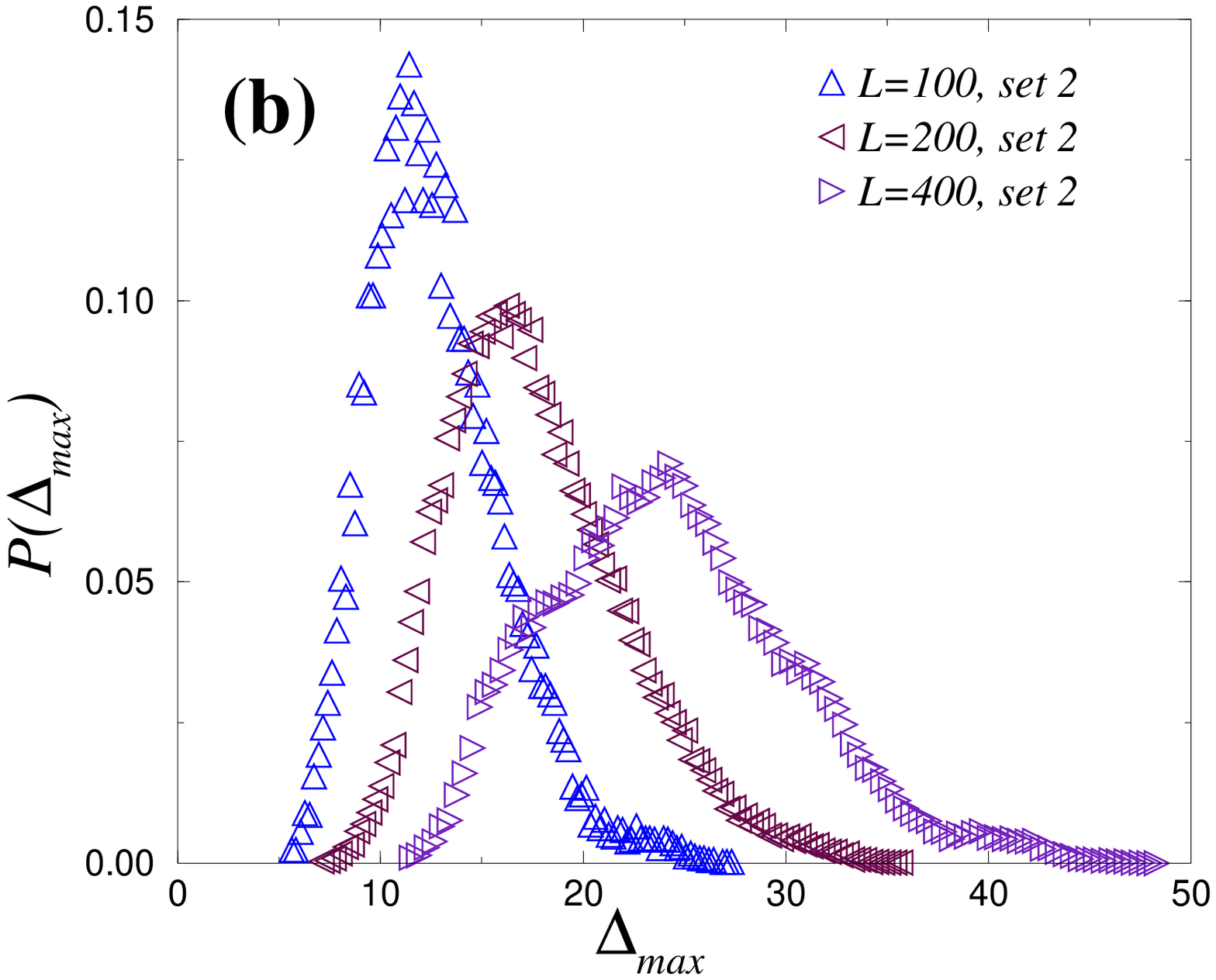}
\vspace*{6.00truecm}
       \includegraphics{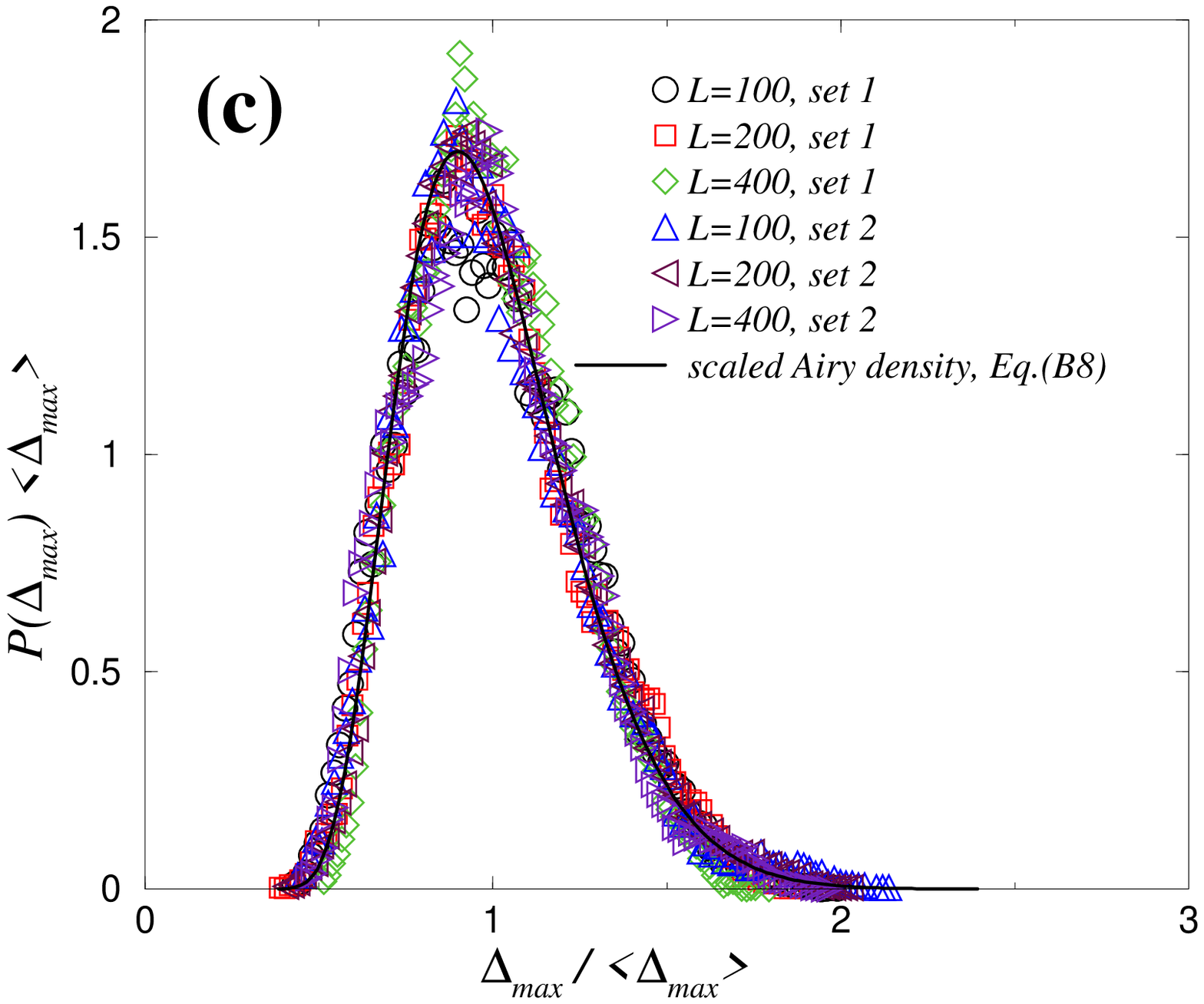}
       \includegraphics{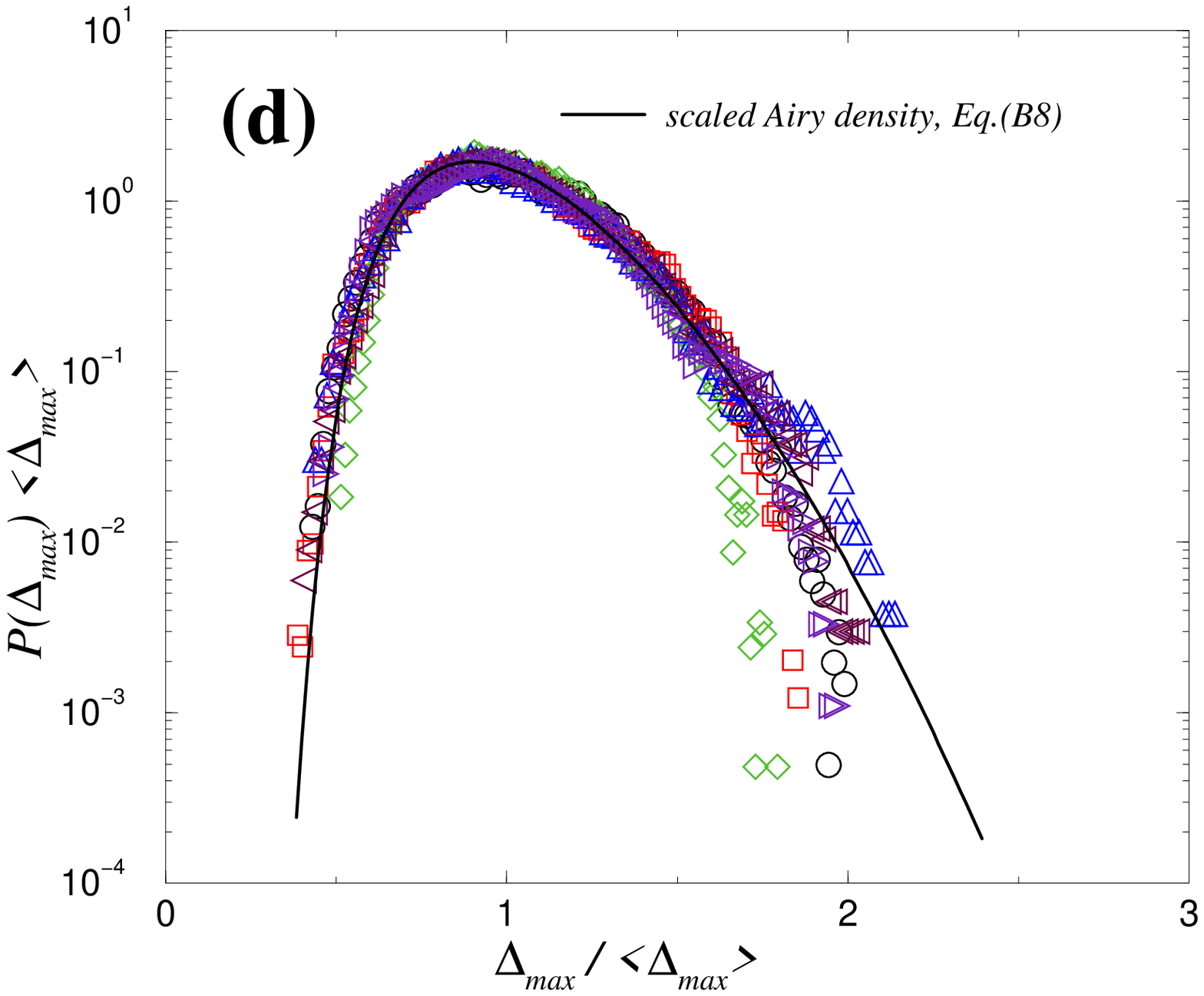}
\vspace*{3.2truecm}
\caption{\small
(a) Steady-state distributions
of the extreme advance, relative to mean front, for several habitat sizes.  Parameters
are $\alpha_1 = 0.5$, $\alpha_2 = 0.7$, and $\mu = 0.2$ (``Set 1''), and
(b) for $\alpha_1 = 0.7$, $\alpha_2 = 0.8$, and $\mu = 0.1$ (``Set
2'').
(c) Scaled histograms for all habitat sizes, and for both Set
1 and Set 2. The solid curve is the scaled Airy density, Eq.~(\ref{P_ext_scaled}) \cite{Majumdar_2004}.
(d) Same data as in (c) but on lin-log scales.}
\label{fig8}
\end{figure}

Figure~\ref{fig8}(a,b) provides an example, using two
different sets of parameters. As habitat size increases, the
mode and dispersion of the estimated probability densities $P(\Delta_{max},L)$
increase, driven by the dependence of $\langle\Delta_{max}\rangle$ on
$L$.  Rescaling these positively skewed histograms, and plotting the resulting
distributions of $(\Delta_{max}/\langle\Delta_{max}\rangle)$, collapses the
data articulately for all system sizes and both sets of parameters onto the
(theoretical) scaled Airy density [Figure~\ref{fig8}(c,d)].
Note that no fits are used in collapsing the scaled data with the
theoretical curve. The largest deviations from the theoretical
prediction can be observed around the peak of the distributions for
some larger system sizes. This poorer statistical sampling is, in
part, due to the correlated nature of the steady-state time series
of the extremes, which is progressively more significant for larger
system sizes.

Longer fronts increase the expected maximal difference between
locations of the front runner and the front's mean position.  The
scaling of the front runner's relative position is possible because
the distribution of the extremes is governed by this single scaling
$\langle\Delta_{max}\rangle\sim L^{1/2}$. Note that traditional
extreme-value limit theorems \cite{Gumbel_1958,Fisher_1928,Berman_1964}
for weakly- or un-correlated fluctuations along the front would yield only a weak logarithmic increase with the
habitat size $L$ \cite{Raychaudhuri_2001,Guclu_2004}.

\begin{figure}[t]
\centering \vspace*{2.50truecm}
       \includegraphics{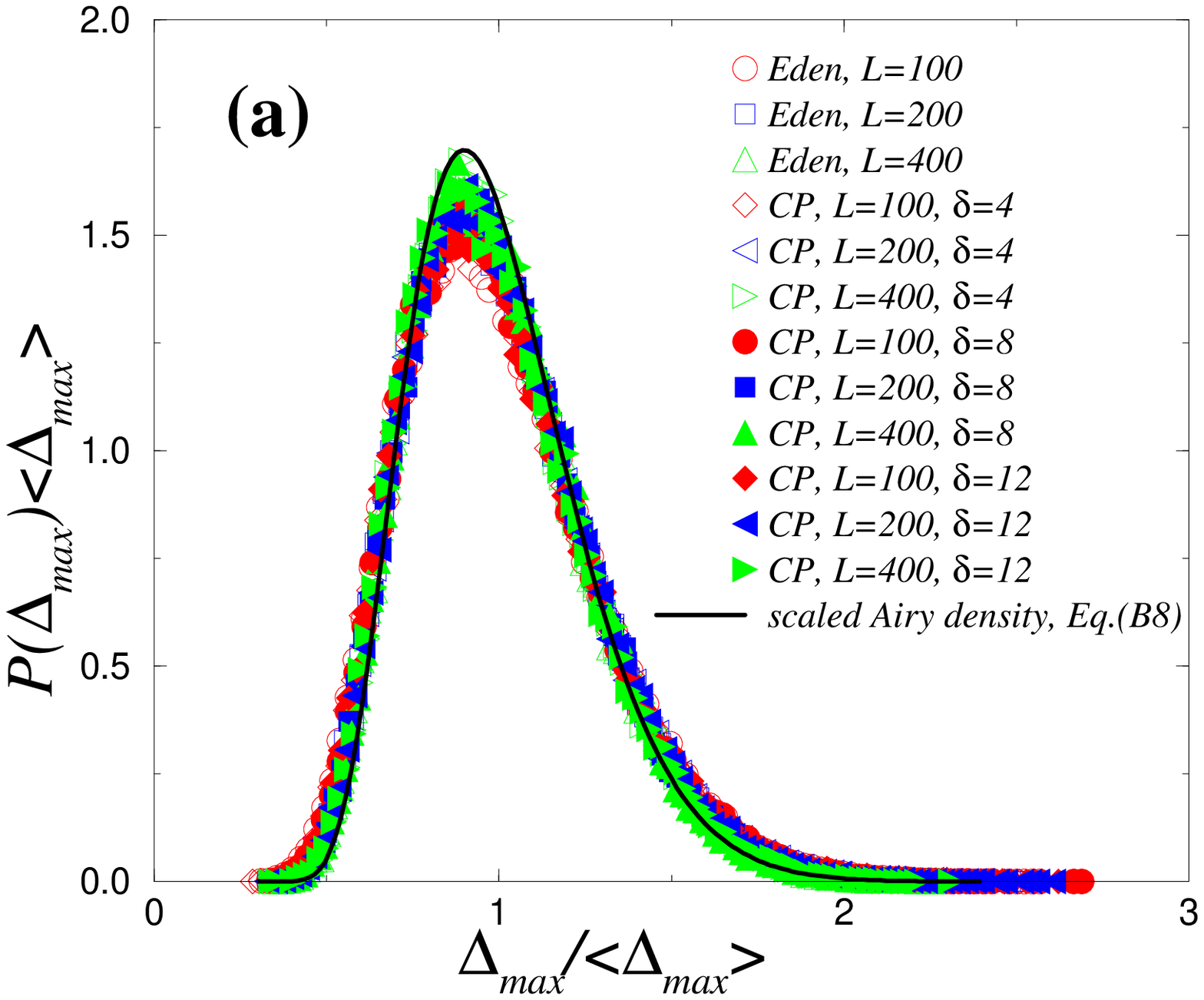}
       \includegraphics{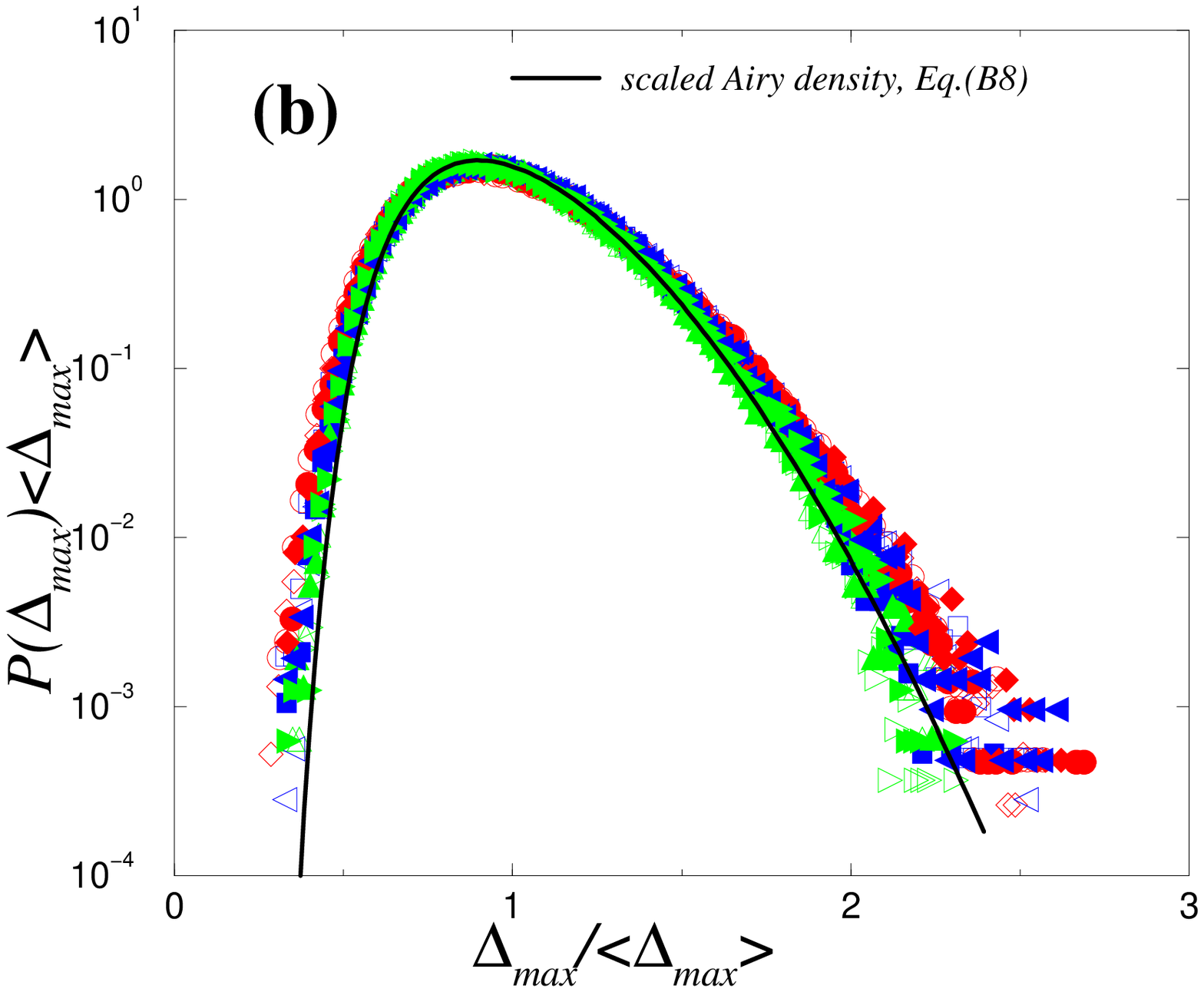}
\vspace*{3.2truecm}
\caption{\small (a) Scaled steady-state
distributions of the extreme advance for the contact process (CP) and for the Eden model
for $L=100, 200, 400$.
For both models there is only one species present (invaders) with local propagation rate
$\alpha_2$ and mortality rate $\mu$.
For the CP, we set $\alpha_2$$=$$1$ and $\mu$$=$$0.20$. The
front dynamics of the CP for $\alpha_2$$=$$1$ and $\mu$$=$$0$ is equivalent
to a variant of the Eden model \cite{Jullien_PRL1985,Jullien_JPA1985},
and we used these parameter values to obtain the Eden data.
For the Eden model, we used only nearest-neighbor propagation ($\delta$$=$$4$).
For the CP, we also explored different neighborhood sizes,
$\delta=4, 8, 12$; all are shown on the plot.
The solid curve is the scaled Airy density, Eq.~(\ref{P_ext_scaled}) \cite{Majumdar_2004}.
(b) Same data as in (a) but on lin-log scales.}
\label{fig9}
\end{figure}

To emphasize the significance of universality, we repeated some of our analyses for
the extremes of the front and compared our model's behavior with with two other, well-known
invasion models, the Eden model and the basic contact process.  We note that the fronts in both of these
models have long been believed to belong to the KPZ class
\cite{Jullien_PRL1985,Jullien_JPA1985,Plischke_1985,Kertesz_1988,Moro_2001,Ferreira_2006}.
The scaled probability densities exhibit the expected, progressive
data collapse on to the universal Airy distribution
[Fig.~\ref{fig9}].
The scaled extreme-distribution plots
include not only different models and system sizes, but for
illustration, different neighborhood sizes as well ($\delta=4, 8,
12$ for the contact process).  The Eden model, the basic contact process,
and our model of invasion under preemptive competition share scaling
properties, revealing common, key properties of their advancing fronts, in the
light of KPZ universality.  Consequently, we can infer a series of ecological
predictions shared by the most often cited discrete, stochastic models of invasive growth.

\section{Discussion}
We recently applied theory for nucleation in homogeneous environments
to the spatial dynamics of invasion \cite{Korniss_JTB2005,OMalley_SPIE2005,OMalley_TPB2006}.
Immigration introduces individuals of an initially rare, but
competitively superior, invader randomly independently in space and
time.  An introduced invader, stochastically, may die or may
propagate locally and initiate a cluster of invaders.  A nucleation
event occurs when an invader cluster reaches a critical size where
further growth through propagation and decline through mortality
have equal probability \cite{Gandhi_1999,Yasi_2006,Allstadt_2007}.
Supercritical clusters, on average, exhibit radial symmetry and expand at
a constant velocity after sufficiently long periods; their growth excludes
the resident competitively.
Circular fronts will reach the same final velocity as
their linear counterparts, with the same corrections as times increases
(O'Malley et al. in press), so that our analysis of linear fronts
furthers application of nucleation theory to the evolutionary ecology of invasion.

To characterize important consequences of discreteness and
stochasticity for invasive advance, we introduced kinetic roughening
of the invader-resident interface.  Roughening of our simulated fronts
behaves as a member of the KPZ-universality class in one
(transverse) dimension. Since steady-state roughening depends on
habitat size, the distribution of the invader's maximal incursion,
relative to the front's mean position, also depends on habitat size.
The advancing front's steady-state velocity correction varies
inversely with habitat size, again a consequence of the
two-dimensional stochastic front's roughening.

Kawasaki et al. (2006) report two-dimensional stochastic simulations
of an invader advancing into empty space; propagation into an open site can occur if any of its 8 nearest
neighbors is occupied.  They also assume that invaders are immortal.
These assumptions, with growth initiated from a compact region occupied by the invaders,
render their model essentially a variant of the Eden model \cite{Jullien_PRL1985}.
Hence their model should exhibit KPZ roughening
\cite{Jullien_JPA1985,Plischke_1985,Kertesz_1988,Ferreira_2006}.

Kawasaki et al. (2006) suggest that stochasticity accelerates
invasive advance in their two-dimensional simulations, and that
reports of deceleration due to stochasticity
may arise because the latter address only one-dimensional
processes \cite{Snyder_2003}.  Above, we detailed simulation results finding that stochasticity
(combined with discreteness) significantly slows the velocity of an
invading competitor in two-dimensions.
Kawasaki et al.'s (2006) seemingly atypical result reflects the
particular deterministic model they compared to their
simulations, rather than dimensionality.  We compared our
stochastic invasion velocities to results from the corresponding
reaction-diffusion model, the common baseline approach to invasion
velocity in ecology \cite{Andow_1990,Holmes_1994}.  In the context of
Moro's (2003) analysis, our
stochastic model has $1 < N^* < 10$; i.e. less than 10 individuals
in each row of the interface width.  Our deterministic comparison,
the reaction-diffusion model, has $N^* \rightarrow \infty$.  Our
stochastic velocity is slower due to diffusion limitation; our
pushed fronts are slower than a pulled front.  Kawasaki et al.
(2006) base their claim on comparing stochastic simulations to
simultaneous advance of the invader in each row.  The invader
advances (periodically) as an exactly straight ``cliff;'' no
roughening can occur.  Their deterministic comparison behaves as if
$N^* \rightarrow 0$, since the front has no width; that is, Kawasaki
et al. (2006) compare their stochastic simulations to an idealized
pushed front, rather than to a pulled front
(both being inadequate to estimate the velocity of strongly-correlated rough fronts).

Antonovics et al. (2006) present a ``patch model'' where local demes
of a single species are coupled by movement between neighboring
locations.  A gradient in density-independent mortality
restricts the species' range, and the density profile predicted
by a mean field model reflects the spatial pattern in mortality.
Stochastic simulations with small deme sizes produce average
density profiles falling below the mean field prediction;
increased rates of diffusion diminish effects of stochasticity,
and density profiles approach the mean field result \cite{Antonovics_2006}.
Hence, a reduction in diffusion limitation may accelerate an advancing
front or extend a static range boundary.

Our model assumes local propagation only, to simulate limited dispersal.
Density profiles at the advancing front strongly constrain mixing of
the invader and resident along the interface. The resulting
diffusion limitation \cite{Moro_2001} distinguishes reaction-diffusion
velocity from invasion velocities driven strictly by local births,
budding, or vegetative propagation.  The very narrow region with
a non-zero density gradient in any row (along the direction of
the invader's advance) of our model's interface indicates little mixing of the
two species, and compares well with Holway's (1998) study of the
invasive Argentine ant. New colonies lie close to existing colonies.
Budding of the invader colonies pushes native ants back as the front
advances, and the width of the front between invader and residents
apparently remains quite narrow.

Although our model has no explicit diffusion, the coarse-graining
procedure does generate, albeit nonlinear, diffusion in the
continuum limit (see Appendix A). Including further local diffusive
moves in our model will not change its
universality class, provided that the effective diffusion
coefficient is not much larger than the effective reaction rate
\cite{Moro_2001}. Thus, one way to change the
asymptotic scaling and the corresponding universality class of the
front is to introduce local diffusion with large rates
(compared to local propagation), or alternatively, introduce
long-range dispersal. In these cases, one expects a crossover from
the KPZ class to that of the deterministic reaction-diffusion
equation, or similar model \cite{Kot_1996}, which exhibits
a pulled front.

The dynamical crossover from pushed to pulled fronts may help
organize results on the modelling and measurement of invasive growth. Generally,
we can associate different fronts with ecological-invasion studies
categorized according to the particular spatial scale emphasized.  At the
within-habitat scale, some plant and animal invaders' spatial growth is
driven by neighborhood-level propagation \cite{DAntonio_1993,Silvertown_1994,Schwinning_1996,Holway_1998}.
The limitation on empirically observed dispersal distances appears consistent with effective-diffusion limitation
and pushed fronts, within any habitat invaded successfully.  Jump dispersal, at the
between-habitat scale, for these same species then must depend on other ecological or
cultural processes generating introduction events \cite{Korniss_JTB2005}.

As estimated dispersal distances of individuals increase \cite{Andow_1990,Bjornstad_2002},
diffusion limitation relaxes, and a reaction-diffusion dynamics may
approximate spatial advance reasonably well, at both the within-habitat
\cite{Dwyer_1992,Frantzen_2000} and between-habitat \cite{vdBosch_1992,Nash_1995}
scales.  Reaction-diffusion models, as noted above, predict pulled fronts
advancing at a constant asymptotic velocity.  Another set
of invasion models can admit increasing velocities; they emphasize
the between-habitat spatial scale and note the importance of long-distance dispersal
for some biological invasions \cite{Kot_1996,Cannas_2006}.
Particular applications include the biogeographical range
expansion of tree species, as indicated by paleoecological records \cite{Clark_1998},
aerial spread of pathogens at the continental scale \cite{Aylor_2003}, and
dispersal vectored by human migratory \cite{Fisher_2001} or economic \cite{Ruiz_2000} activity.
Long-distance dispersal events, even if rare, can establish colonies detached
from the main body of invaders; the two populations may then grow and coalesce
\cite{Shigesada_1995}. Invasion dynamics averaging over dispersal-distance distributions
associated with these models can predict pulled fronts with an ever-increasing spatial velocity;
Clark et al. (2001) discuss details and theoretical modifications.

\subsection{The front-runner}
Our front-runner analyses integrate a series of stochastic
spatial models through a few universal scaling relationships.
For all models belonging to the KPZ universality class (in one transverse dimension) and
for large enough habitat size $L$, $\langle\Delta_{max}\rangle$$\sim$$L^{1/2}$ [Fig.~\ref{fig3}(b)];
the distribution of the scaled probability density is given by the
Airy distribution [Figs.~\ref{fig8} and \ref{fig9}].
These universal features will be exhibited
by any invasion model belonging to the KPZ class
\cite{Plischke_1985,Ferreira_2006}, not just
our two-species model, the contact process,
and the Eden model.

We referred to transverse system size $L_y=L$ as habitat size, not
ordinarily a ``tuneable parameter" for probing roughness
characteristics of the interface. Furthermore, for simplicity, we
employed periodic boundary conditions in the transverse direction.
From the viewpoint of empirical application, we can envision a
scenario where one has access to only a limited observation window
on the full landscape.  Designate the window's size $L_{\rm obs}$, where $1\ll L_{\rm
obs}\ll L$. These length scales are measured in units of the minimum
linear size needed to sustain a single individual (linear size of a lattice site in simulations).
One of the research challenges in extreme-value statistics is to predict some
characteristic of the extremes on global scales (i.e., at the entire
habitat or landscape level) based on some limited local measurements
(i.e., taken from the observation window). One can divide the
observation region into a number of smaller non-overlapping
subwindows of size $l$, such that $1\ll l\ll L_{\rm obs}$, and by
changing the size of the subwindow $l$, perform a finite-size
analysis to extract estimates for the roughness exponent from the
average width and the extremes as $w(l)\sim l^{\alpha}$ and
$\langle\Delta_{max}(l)\rangle\sim l^{\alpha} $. The subwindows will
now have ``window" or free boundary conditions
\cite{Antal_2001,Antal_2002,Majumdar_2005}, clearly differing from
the periodic boundaries we simulated. However, boundary conditions
do not affect the roughness exponent, hence the scaling properties
of width and that of the extremes with $l$. Therefore, one can
estimate the size of the extremes for the full landscape of size $L$
from measurements obtained in a limited observation window,
$\langle\Delta_{max}(L)\rangle \approx \langle\Delta_{max}(L_{\rm obs})\rangle (L/L_{\rm obs})^{\alpha}$.
In contrast, were the interfacial fluctuations weakly- or
un-correlated, traditional extreme-value statistics
(for short-tailed parent distributions) would apply. In turn, extrapolating
local measurements to global scales for the extremes would yield
only marginal (logarithmic) increase,
$\langle\Delta_{max}(L)\rangle \approx \langle\Delta_{max}(L_{\rm
obs})\rangle [\ln(L)/\ln(L_{\rm obs})]^{1/\gamma}$,
where $\gamma$ is a non-universal exponent, and is related the
precise shape of the exponential-like tail of the parent densities
\cite{Raychaudhuri_2001,Guclu_2004}.

The velocity of spatial advance and the properties of the
interface region remain objects of study in population biology
because of their conceptual significance and potential importance
to applied ecology. Despite attention
directed to invasive spatial growth, ecological theory has
overlooked the impact of correlations along an advancing front
(roughening in a two-dimensional environment). Although these
spatial correlations arise through local interactions, at a scale
sometimes considered ``noise,'' they can directly affect
population/community-level phenomena: invader velocity, dynamics at
the interface, and relative position of the front-runner.

\section*{Acknowledgments}
We thank B. Kozma for discussions, earlier collaborations
\cite{OMalley_PRE2006,OMalley_SPRINGER}, and permission to use
his data in this work. We are also grateful to S. Majumdar for
providing us the numerically evaluated Airy distribution
\cite{Majumdar_2004,Majumdar_2005} shown in Figs.~\ref{fig8} and
\ref{fig9}. E. Moro kindly commented on earlier
results, and we also appreciate discussion with A. Allstadt
and  Z. R\'{a}cz. We also thank the reviewer for offering
useful suggestions.  This work was supported in part by the National
Science Foundation under Grant Nos. DEB-0324689 and DMR-0426488.

\appendix
\section*{Appendix}

\section{Deterministic reaction-diffusion limit and pulled fronts}
\renewcommand{\theequation}{A.\arabic{equation}} 
\setcounter{equation}{0}  
O'Malley et al. (2006b) write the master equation corresponding to
transition rates in Expression~(\ref{rates}) \cite{vKampen_1981}.
Assuming that densities at different sites remain uncorrelated
\cite{McKane_2004,Korniss_1997}, one obtains the dynamics of the
ensemble-averaged local densities $\rho_i({\bf
x},t)$$\equiv$$\langle n_i({\bf x},t)\rangle$,
\begin{eqnarray}
\rho_i({\bf x},t+1) - \rho_i({\bf x},t) & = & \left[1 -
\rho_1({\bf x},t) - \rho_2({\bf x},t)\right]
\nonumber \\
\times \frac{\alpha_i}{4}\sum_{{\bf x'}\epsilon {\rm nn}({\bf
x})}\rho_{i}({\bf x'},t)
& - & \mu \rho_i({\bf x}, t) \;,
\label{lattice_RD}
\end{eqnarray}
where $i = 1, 2$ for resident and invader species, respectively, and we let $\delta$$=$$4$ for simplicity.
Taking the naive continuum limit of the above equations yields the (coarse-grained) equations of motion
\begin{eqnarray}
\frac{\partial \rho_i({\bf x},t)}{\partial t} & = &
\frac{\alpha_i}{4} \left[1 - \rho_1({\bf x},t) - \rho_2({\bf
x},t)\right] \nabla^2\rho_i({\bf x},t)
\nonumber\\
& + &  \alpha_i \left[1 - \rho_1({\bf x},t) - \rho_2({\bf
x},t)\right] \rho_i({\bf x},t)\nonumber \\
& - & \mu \rho_i({\bf x},t)\;,
\label{continuum_RD}
\end{eqnarray}
$i=1,2$, and $\nabla^2$ is the Laplacian operator. In statistical physics, equations of this
kind are referred to as ``mean-field" theory, in the sense that correlations
between spatial degrees of freedom are neglected \cite{McKane_2004,Antonovics_2006}.
This contrasts to population biology's terminology
where ``mean-field" ordinarily refers to non-spatial
models. After adding appropriate noise terms to the above equations, the
resulting Langevin equation can, in principle, effectively capture
both spatial and stochastic effects, identical to those in the
underlying discrete individual-based model
\cite{Schmittmann_1995,Hinrichsen_2000,vKampen_1981,Gardiner_1985}.

The spatially homogeneous solutions of the above  equations,
$(\rho_1^*,\rho_2^*)$, are $(0,0), (1 - \mu/\alpha_1,0)$, and $(0,1
- \mu/\alpha_2)$; coexistence is not feasible. In the parameter
regime of interest, $\mu < \alpha_1 < \alpha_2$, only the last
solution $(0,1 - \mu/\alpha_2)$ is stable.  The advance of a front
separating the stable $(0,1 -\mu/\alpha_2)$ (invader dominated) and
unstable $(1 - \mu/\alpha_1,0)$ (resident dominated) regions amounts
to propagation into an unstable state
\cite{Fisher_1937,Kolmogorov_1937,Aronson_1978}, a phenomenon that
has generated a vast amount of literature \cite{vSaarloos_2003}.
Given the mean-field equations~(\ref{continuum_RD}), the front is ``pulled'' by the
leading edge into the unstable state. For a sufficiently sharp
initial density profile \cite{Murray_2003}, the asymptotic velocity
is determined by the infinitesimally small density of invaders that
intrude into the linearly unstable region dominated by the resident
species. Linearizing Eqs.~(\ref{continuum_RD}) about the unstable state,
$\rho_1=1-\mu/\alpha_1 + \phi_1$, $\rho_2=\phi_2$, one obtains for
the density of invaders
\begin{equation}
\frac{\partial \phi_2({\bf x},t)}{\partial t} =
\frac{\mu}{4}\frac{\alpha_2}{\alpha_1}\nabla^2\phi_2({\bf x},t)
+ \mu\left(\frac{\alpha_2}{\alpha_1} - 1\right)\phi_2({\bf x},t) \;.
\label{effective_RD}
\end{equation}
Performing standard analysis \cite{Murray_2003,vSaarloos_2003} of
this equation with $D\equiv(\mu/4)(\alpha_2/\alpha_1)$ and
$r\equiv\mu(\alpha_2/\alpha_1-1)$, the effective diffusion and
reaction coefficients, respectively, we obtain the asymptotic
velocity of the ``marginally stable" invading fronts,
$v_p=2\sqrt{Dr}$, i.e.,
\begin{equation}
v_{p} = \frac{\mu}{\alpha_1} \sqrt{\alpha_2 (\alpha_2 - \alpha_1)}.
\label{velocity_pulled}
\end{equation}
$v_{p}$ is the minimal velocity of a traveling wave permitted by
Eqs.~(\ref{continuum_RD}), and is actually realized by the
deterministic nonlinear reaction-diffusion dynamics for sufficiently
sharp initial profiles \cite{Murray_2003,vSaarloos_2003}. Note that
we also performed the numerical iteration of the nonlinear
difference equations Eqs.~(\ref{lattice_RD}) [the underlying natural
discretization of the corresponding reaction-diffusion system
Eqs.~(\ref{continuum_RD})] \cite{OMalley_PRE2006,OMalley_SPRINGER}.
The results show strong agreement with velocity of the linearized
equations Eq.~(\ref{velocity_pulled}) [Fig.~\ref{fig4}(b)]. Despite the nonstandard
nonlinear (but deterministic) system of Eqs.~(\ref{continuum_RD}), the
front velocity can be fully governed by the linearized system where the
front is pulled by the leading edge.

In discrete individual-based models with limited diffusion
(resulting in strongly correlated front fluctuations), one does not
expect Eq.~(\ref{velocity_pulled}) to provide a good approximation
for the asymptotic front velocity \cite{Moro_2001,Moro_2003}.
Instead, Eq.~(\ref{velocity_pulled}) can serve as a baseline
reference for comparison with the actual front velocity. For the
deterministic front just discussed, where invasion is dominated by
infinitesimal densities at the leading edge, one can also obtain the
width of the interfacial region \cite{Murray_2003,vSaarloos_2003},
$w\simeq\sqrt{D/r}=(1/2)(1-\alpha_1/\alpha_2)^{-1/2}$, independently
of the landscape size $L_y$.

In passing, we consider the velocity results for our model's reaction-diffusion
approximation in light of a spatial-competition model by Lewis et al. (2002).
Our model assumes preemptive competition.  Since resident and invader use the limiting
resource so similarly (interact strongly), our model's mean-field approximation
prohibits coexistence.  The invader's advance translates one boundary equilibrium
into another.  As noted above, iterating Eqs.~(\ref{lattice_RD}) produces a velocity
matching the pulled-front velocity $v_p$ predicted by the linearizing our model's
reaction-diffusion approximation.  That is, for all parameter combinations we
evaluated, the ``linear conjecture'' \cite{Mollison_1995} predicts the mean-field velocity.

Lewis et al. (2002) write a two-species competition model of Lotka-Volterra form,
with each species diffusing along a one-dimensional continuum.  The spatially homogeneous
equilibria are standard; coexistence is stable if self-regulation in both species is stronger
than interspecific competition.  The authors derive a set of
sufficient conditions under which the linear conjecture will predict their
model's invasion velocity.  Outside parameter ranges specified by these conditions,
their model's velocity might or might not match that obtained by linearizing
the dynamics at the invasive front's leading edge.  Put simply, they find that the
linear conjecture holds when both diffusion of the competitively superior invader
is not too small compared to the resident's diffusion, and the species do
not interact too strongly.  Our results, Fig.~(\ref{fig4}b),
are numerical; if the result for Lotka-Volterra competition with diffusion
\cite{Lewis_2002} applies more generally, our mean-field approximation's
velocity might or might not match $v_p$ so closely for other parameter values.
We emphasize strongly that our
important results on invasion velocity concern the differences between
the discrete, stochastic model's velocity (a pushed front) and that of the
associated reaction-diffusion model (a pulled front), quite independently
of whether or not the latter can be approximated via the linear conjecture.

\section{The KPZ equation and scaling behavior of the KPZ universality class}
\renewcommand{\theequation}{B.\arabic{equation}} 
\setcounter{equation}{0}  
The interface dynamics of a large class of discrete
individual-based models can be described by the effective (or
coarse-grained) Langevin equation \cite{Kardar_1986}
\begin{equation}
\frac{\partial h(y,t)}{\partial t} = \nu\frac{\partial^2 h(y,t)}{\partial y^2} +
\lambda \left(\frac{\partial h(y,t)}{\partial y}\right)^2 + \zeta(y,t) \;
\label{KPZ}
\end{equation}
where $\nu$ is the effective ``surface tension," $\lambda$ is the
nonlinear coupling constant, and $\zeta(y,t)$ is delta-correlated
Gaussian noise, $\langle\zeta(y,t)\rangle=0$ and
$\langle\zeta(y,t)\zeta(y',t')\rangle=2\Gamma\delta(y-y')\delta(t-t')$,
with $\Gamma$ being the noise intensity. $\delta(\ldots)$ is the Dirac
delta. The coupling constants and the noise intensity
depend strongly on the details of the course-graining procedure
(i.e., the local rates or the finite neighborhood size in the original discrete model, and how the
continuum limit was taken from that model). The variable $h(y,t)$ is
the coarse-grained (or mesoscopic) limit of the local front advancements $h_y(t)$.
In some cases, this procedure can be carried out explicitly from
first principles \cite{Plischke_1987,Korniss_2000,Korniss_2003}.
More often, however, due to the complexity of the details of the
model, this task is insurmountable. Even then, however, fundamental
symmetry considerations can strongly motivate the applicability of
the KPZ equation. Thus, equations of this sort
should not be interpreted as a rigorous continuum limit derived from
the local interface rules, but rather as a coarse-grained description. This approach has been enormously
successful in advancing understanding of large-scale morphological properties of
interfaces and surfaces in complex natural and artificial systems
\cite{Barabasi_1995,Halpin_1995,Korniss_2000,Bru_2003}.
A number of models in the large-scale limit
can be effectively described by Eq.~(\ref{KPZ}), i.e., interesting
attributes of these models (e.g., the width, the extreme
advance, and their distributions) exhibit the same scaling
behavior and shape as given by the solution of Eq.~(\ref{KPZ}).
These models are said to be belong to the KPZ universality
class.

\subsection{Temporal and habitat-size scaling of KPZ fronts}

The temporal and habitat-size scaling of the width $\langle
w^2(L,t)\rangle$ typically identifies the universality class of an advancing
front. In finite systems, the width grows as $\langle
w^2(L,t)\rangle$$\sim$$t^{2\beta}$ from early to intermediate times.
At a system-size-dependent crossover time, $t_\times$$\sim$$L^z$, it
saturates (reaches steady state) and scales as $\langle
w^2_{sat}\rangle \equiv\langle w^2(L,\infty)\rangle\sim
L^{2\alpha}$, where $L$ is the transverse linear system size.
$\alpha$, $\beta$, and  $z$ are referred to as the roughness,
growth, and dynamic exponents, respectively.  The exponents obey the scaling relationship
$\alpha=\beta z$. The full temporal and finite-size behavior
(reproducing the growth and the steady state regimes as limiting cases)
can be captured by the Family-Vicsek scaling form \cite{Family_1985}
\begin{equation}
\langle w^2(L,t)\rangle = L^{2\alpha}f(t/L^z).
\label{FV}
\end{equation}
For small values of its argument, $f(x)$ behaves as a power law,
while for large arguments it approaches a constant
\begin{equation}
f(x)=\left\{
\begin{array}{ll}
x^{2\beta}       & \mbox{for $x \ll 1$} \\
{\rm const.}     & \mbox{for $x \gg 1$}
\end{array} \right. \;,
\label{f_x}
\end{equation}
yielding the scaling behavior of the width, first in the growth regime and
then in the steady-state regime [Eq.~(\ref{w2_scale}), Fig.~\ref{roughening}],
provided the scaling relationship $\alpha=\beta z$ holds.
For the KPZ universality class,
these exponents can be obtained exactly, $\alpha=1/2$, $\beta=1/3$,
and $z=3/2$ \cite{Kardar_1986,Barabasi_1995}.
Furthermore, in the steady state, employing path-integral techniques,
both the width distribution \cite{Foltin_1994}
and the distribution of the extremes \cite{Majumdar_2004}
were obtained analytically. Scaling behaviors then can be
compared to those obtained from a discrete model (e.g., by simulations),
indicating whether or not the discrete stochastic model belongs to the KPZ universality class.

\subsection{Steady-state width distribution of the KPZ universality class}
\begin{figure}[t]
\centering
\vspace*{2.50truecm}
       \includegraphics{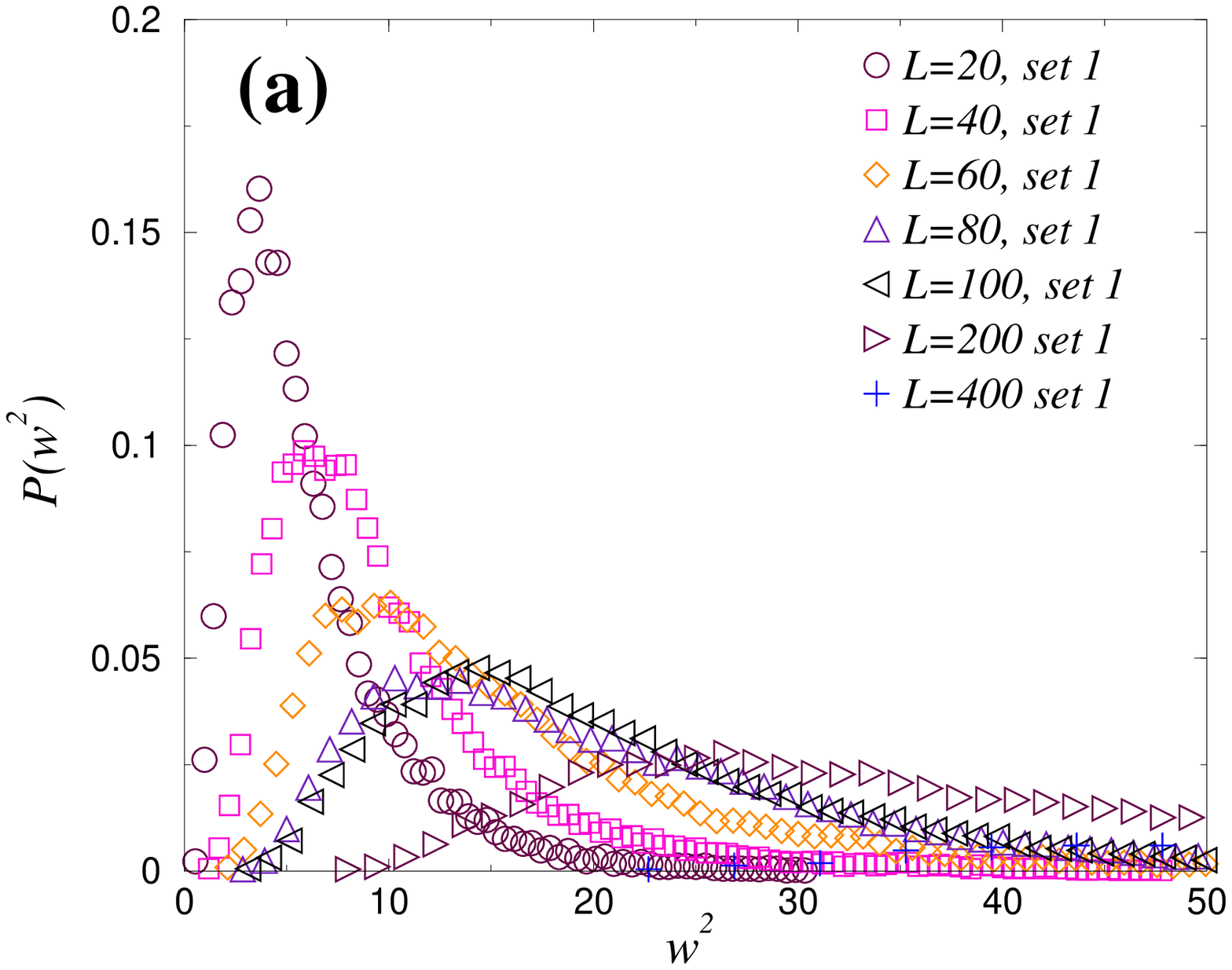}
       \includegraphics{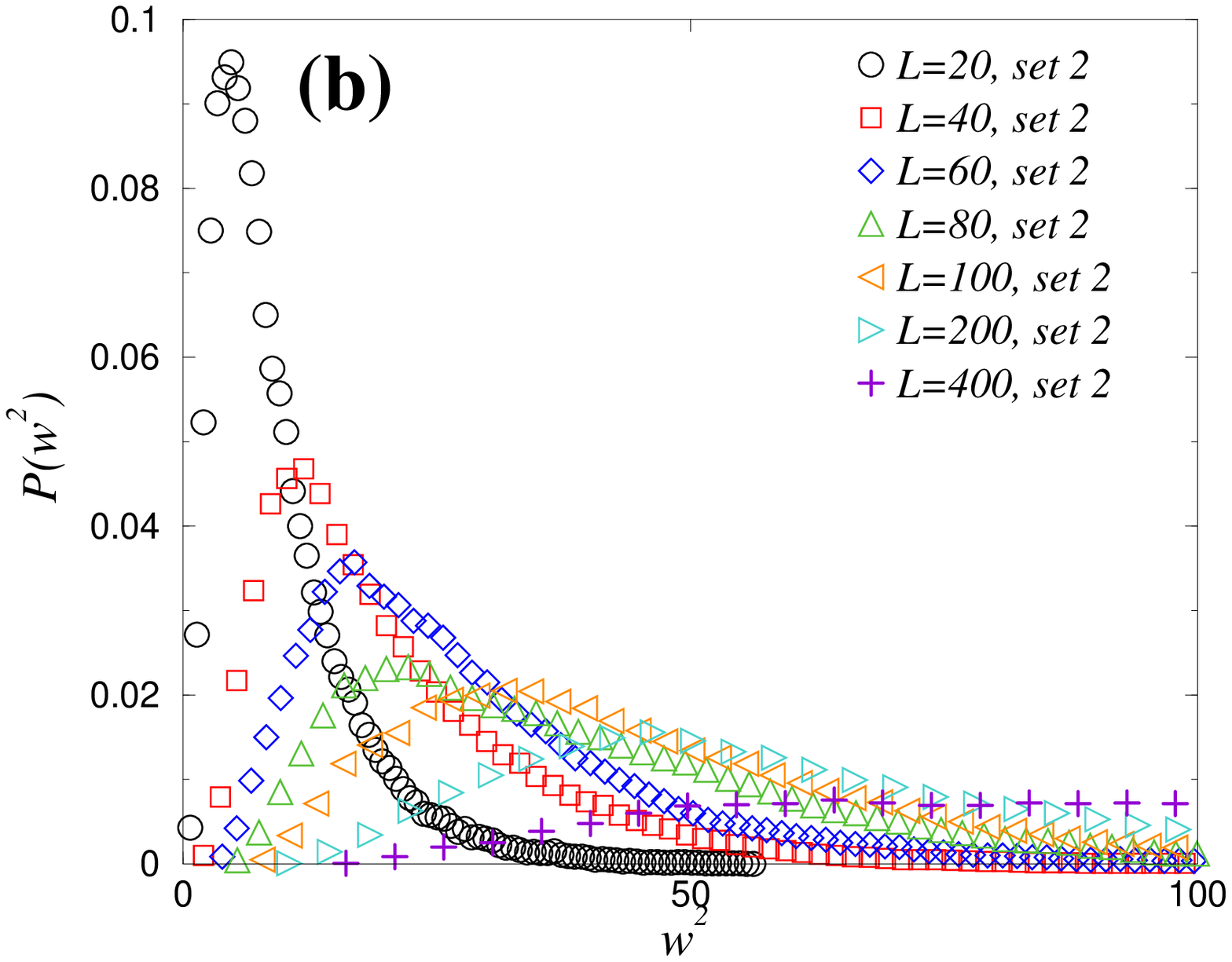}
\vspace*{6.00truecm}
       \includegraphics{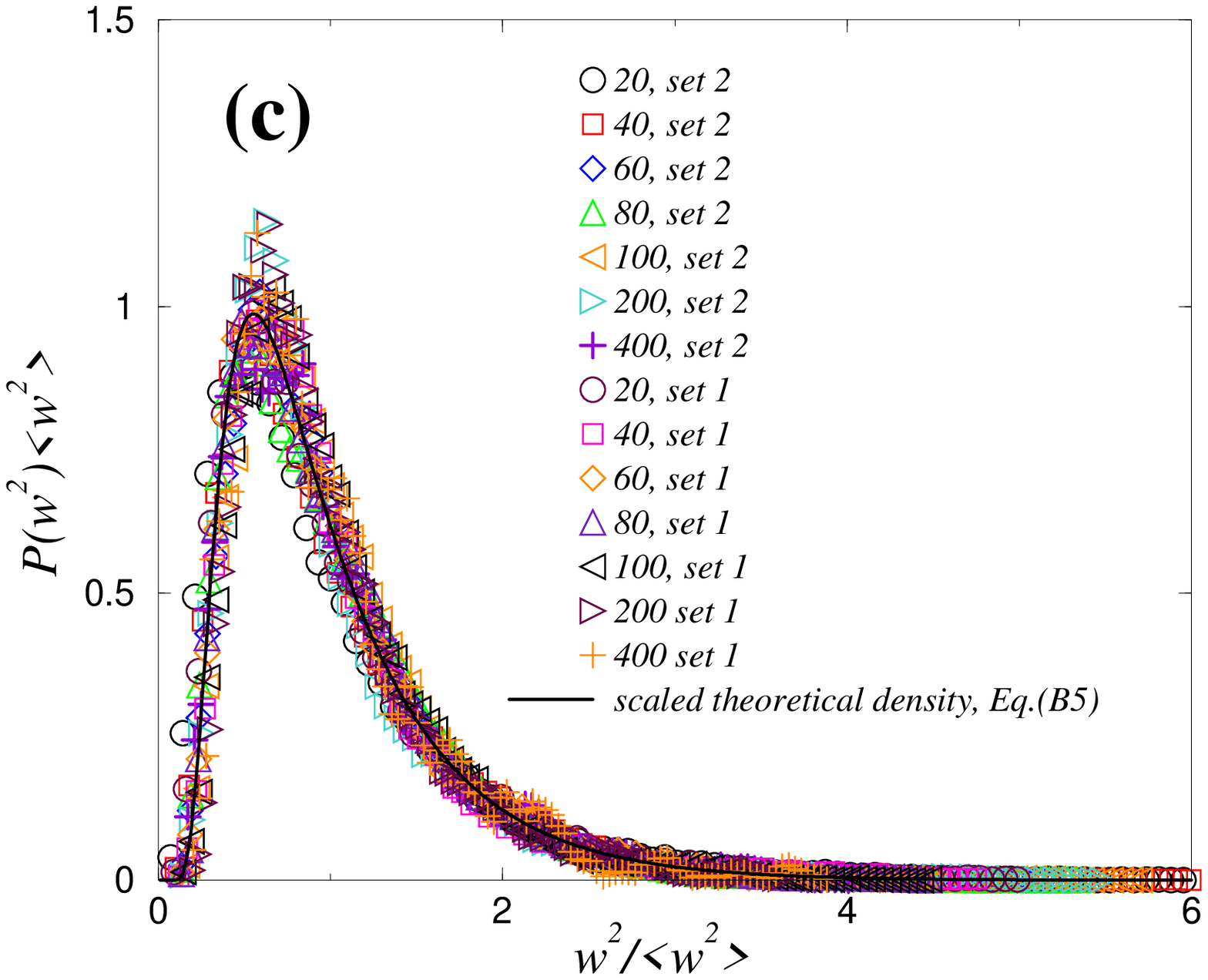}
       \includegraphics{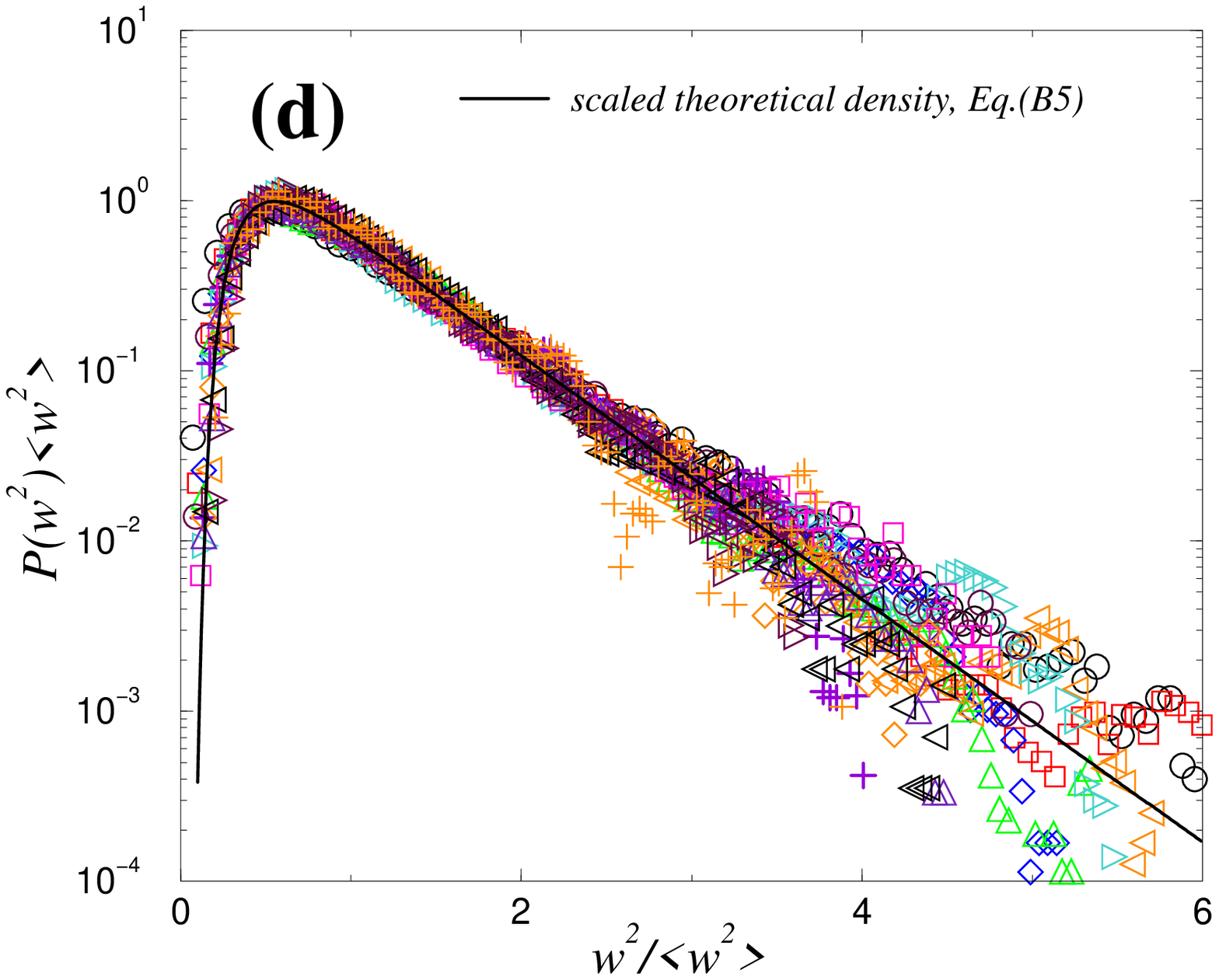}
\vspace*{3.2truecm}
\caption{\small
(a) Steady-state width distributions in our two-species invasive front for several habitat sizes
with parameters $\alpha_1$$=$$0.5$, $\alpha_2$$=$$0.7$, $\mu$$=$$0.2$ (``Set 1''); and
(b) for $\alpha_1$$=$$0.7$, $\alpha_2$$=$$0.8$, $\mu$$=$$0.1$ (``Set 2'').
(c) Scaled histograms for all habitat sizes, for both Set 1 and Set 2.
The solid curve is the scaled analytic width distribution of the KPZ class, Eq.~(\ref{P_w2}) \cite{Foltin_1994}.
(d) Same scaled data as in (c), but on lin-log scales.}
\label{fig5}
\end{figure}

The width distribution typically provides a strong signature of
the underlying universality class of the fluctuating, growing
interface \cite{Antal_2001,Antal_2002}.  It is important to note that for models with local dispersal,
the measured probability density (or histogram) of the individual
(row $y$) interface fluctuations (or ``parent" densities) $P(h_y-\overline{h})$ have tails that
decay faster than any power law (approximately
Gaussian in our model, but this local property depends strongly on model details).
We also note that if the height fluctuations
$(h_y-\overline{h})^2$ in Eq.~(\ref{w2}) were independent (or their
correlation along the interface decayed sufficiently fast), the
probability density of $w^2$ would be Gaussian, by the
central limit theorem. Instead, for our roughened front,
these variables are strongly correlated. As universality
arguments suggest, the shape of the steady-state distribution of the
width for all (sufficiently large) system sizes and for all models
belonging to the KPZ class will be identical, governed by a single
scale.  That is, the average width $\langle w^2\rangle\sim L$. Hence,
the width distribution for any model in this class with rough fronts can be
written as
\begin{equation}
P(w^2,L)=
\frac{1}{\langle w^2\rangle}\Phi(w^2/\langle w^2\rangle)\;,
\label{P_w2_hist}
\end{equation}
where $\Phi(s)$ is the probability density of the scaled variable
$s\equiv w^2/\langle w^2\rangle$. $\Phi(s)$ for periodic boundary conditions (used in our
simulations) was obtained by Foltin et al. (1994), and is given by
\begin{equation}
\Phi(s) = \frac{\pi^2}{3}\sum_{n = 1}^\infty (-1)^{n-1}n^2 e^{-(\pi^2 / 6) n^2 s} \;.
\label{P_w2}
\end{equation}
The universal scaling function above can be employed to test whether
a discrete individual-based invasion model with a rough front
belongs to the KPZ class. Analyzing the width histograms, as
empirical measures of the theoretical probability density, supports
our conclusion that the two-species invasion model, the focus of our
investigations, belongs to the KPZ class [Fig.~\ref{fig5}].
Note that no fits were used in collapsing the scaled data to the theoretical curve.
Analytic scaling functions for the width distribution can also be obtained
for other types of  boundary conditions, e.g., ``window" or free
\cite{Antal_2002}.

\begin{figure}[t]
\centering \vspace*{2.50truecm}
       \includegraphics{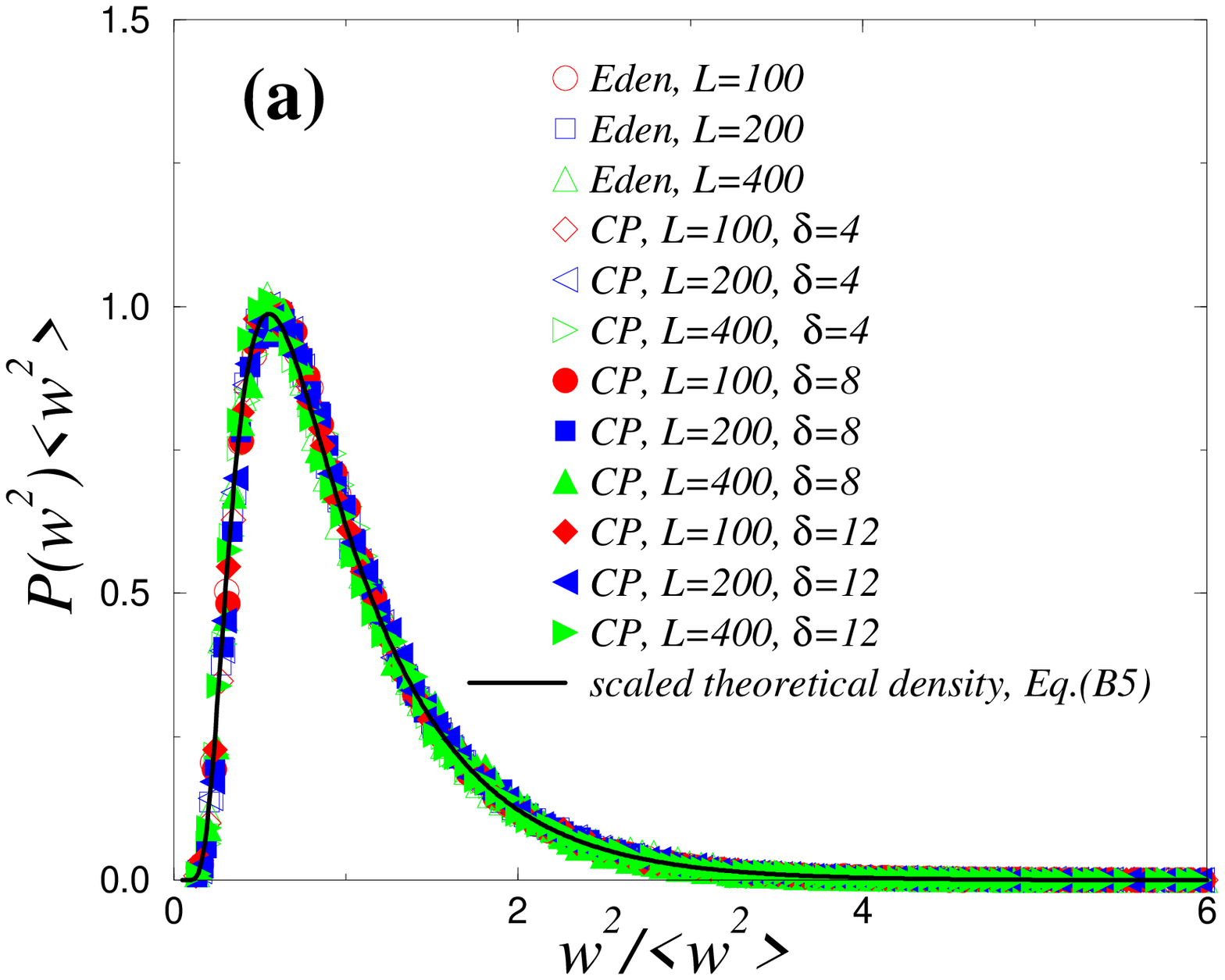}
       \includegraphics{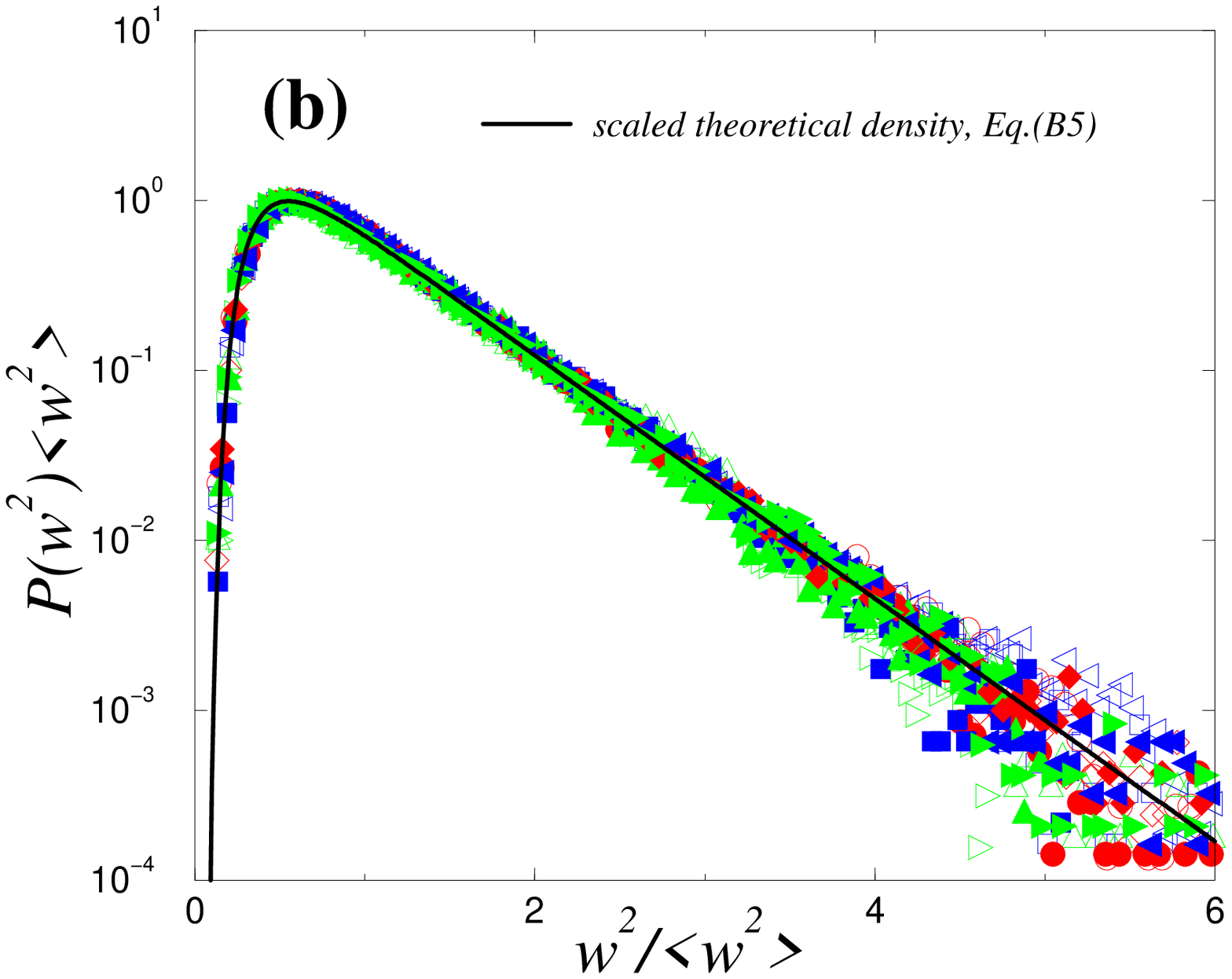}
\vspace*{3.2truecm}
\caption{\small (a) Scaled steady-state width distributions
for the contact process (CP) and for the Eden model for $L=100, 200, 400$.
For both models there is only one species present (invaders) with local propagation rate
$\alpha_2$ and mortality rate $\mu$.
For the CP, we set $\alpha_2$$=$$1$ and $\mu$$=$$0.20$. The
front dynamics of the CP for $\alpha_2$$=$$1$ and $\mu$$=$$0$ is equivalent
to a variant of the Eden model \cite{Jullien_PRL1985},
and we used these parameter values to obtain the Eden data.
For the Eden model, we used only nearest-neighbor propagation ($\delta$$=$$4$).
For the CP, we also explored different neighborhood sizes,
$\delta=4, 8, 12$; all are shown on the plot. The
solid curve is the scaled analytic KPZ width distribution, Eq.~(\ref{P_w2}) \cite{Foltin_1994}.
(b) Same scaled data as in (a) but on lin-log scales.}
\label{fig_Pw2_ECP}
\end{figure}

To emphasize the concept of universality, we also
analyzed data and constructed histograms for propagating fronts in
two well-known invasion models, the Eden model and the contact
process. We note that the fronts in both of these individual-based
models have long been believed to belong to the KPZ class
\cite{Jullien_PRL1985,Jullien_JPA1985,Plischke_1985,Kertesz_1988,Ferreira_2006,Moro_2001}.
The scaled probability densities exhibit progressive data collapse
onto the universal scaled KPZ density [Fig.~\ref{fig_Pw2_ECP}], offering
further support that the invasive fronts in the Eden model and
in the basic contact process belong to the KPZ class.

\subsection{Asymptotic velocity of KPZ fronts: temporal and habitat-size corrections}
As described in the text, a model's
asymptotic front velocity cannot be universal.  However,
Krug and Meakin (1990) showed that the forms of the
temporal and finite-size ($L_{y}=L$) corrections are universal
within a given class. Above we demonstrated that the
propagating interface in our two-species invasion model belongs to
the KPZ universality class with exponents $\alpha=1/2$, $\beta=1/3$,
and $z=3/2$. Using Krug and Meakin's result, one obtains the
corrections to the front velocity $v(t, L)$,
\begin{equation}
v(t,L)= \left \{ \begin{array}{lll}
v^* - c_1 t^{-2(1-\alpha)/z} = v^* - c_1 t^{-2/3} & \mbox{for} & t \ll L^z \\
v^* - c_2 L^{-2(1-\alpha)} = v^* - c_2 L^{-1}   & \mbox{for} & t \gg
L^z
\end{array}\right. \;,
\label{velocity_corr}
\end{equation}
where $c_1$ and $c_2$ are non-universal constants that depend, as does
$v^*$, on propagation and mortality rates at the level of individual
competitors.

\begin{figure}[t]
\centering \vspace*{2.50truecm}
       \includegraphics{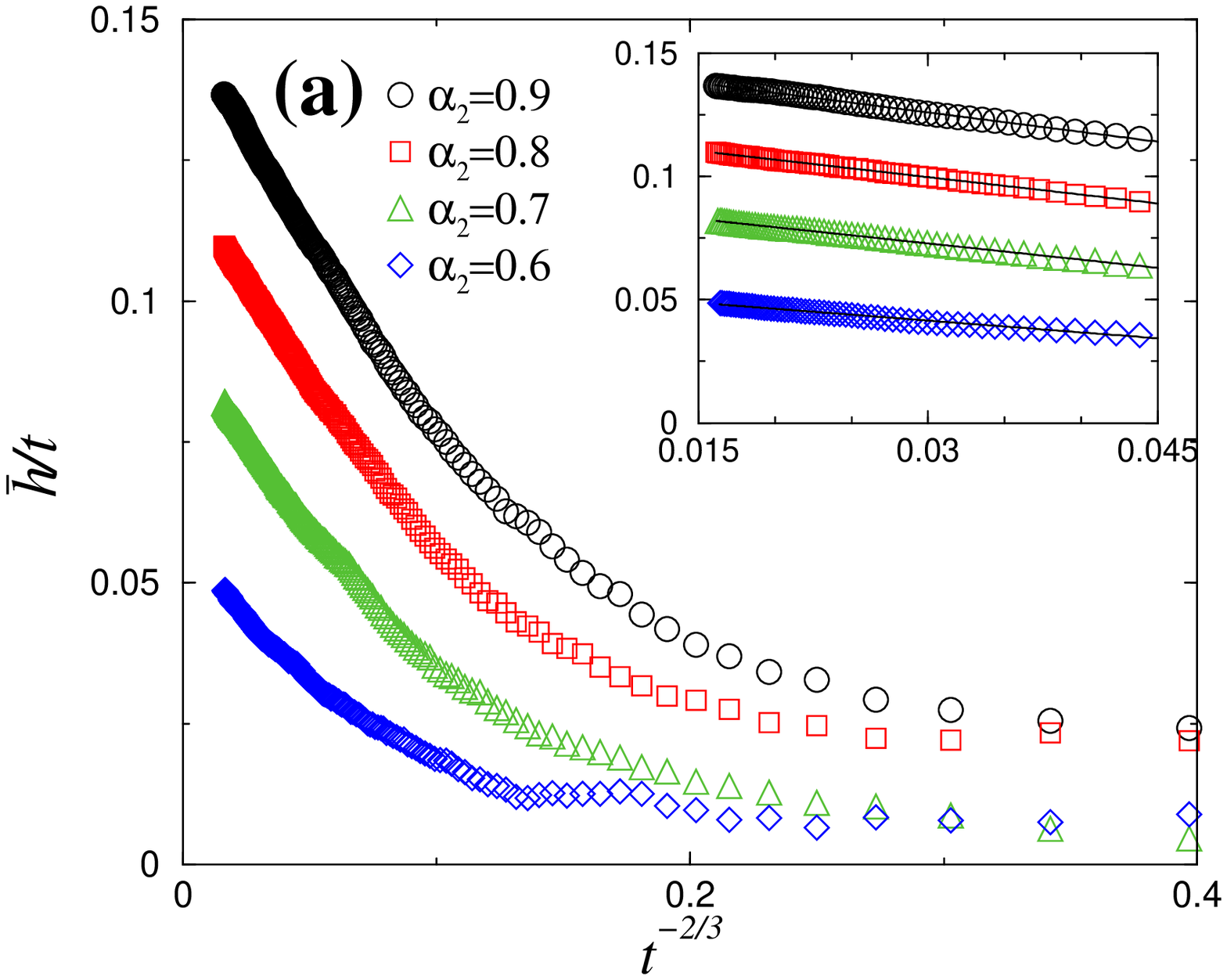}
       \includegraphics{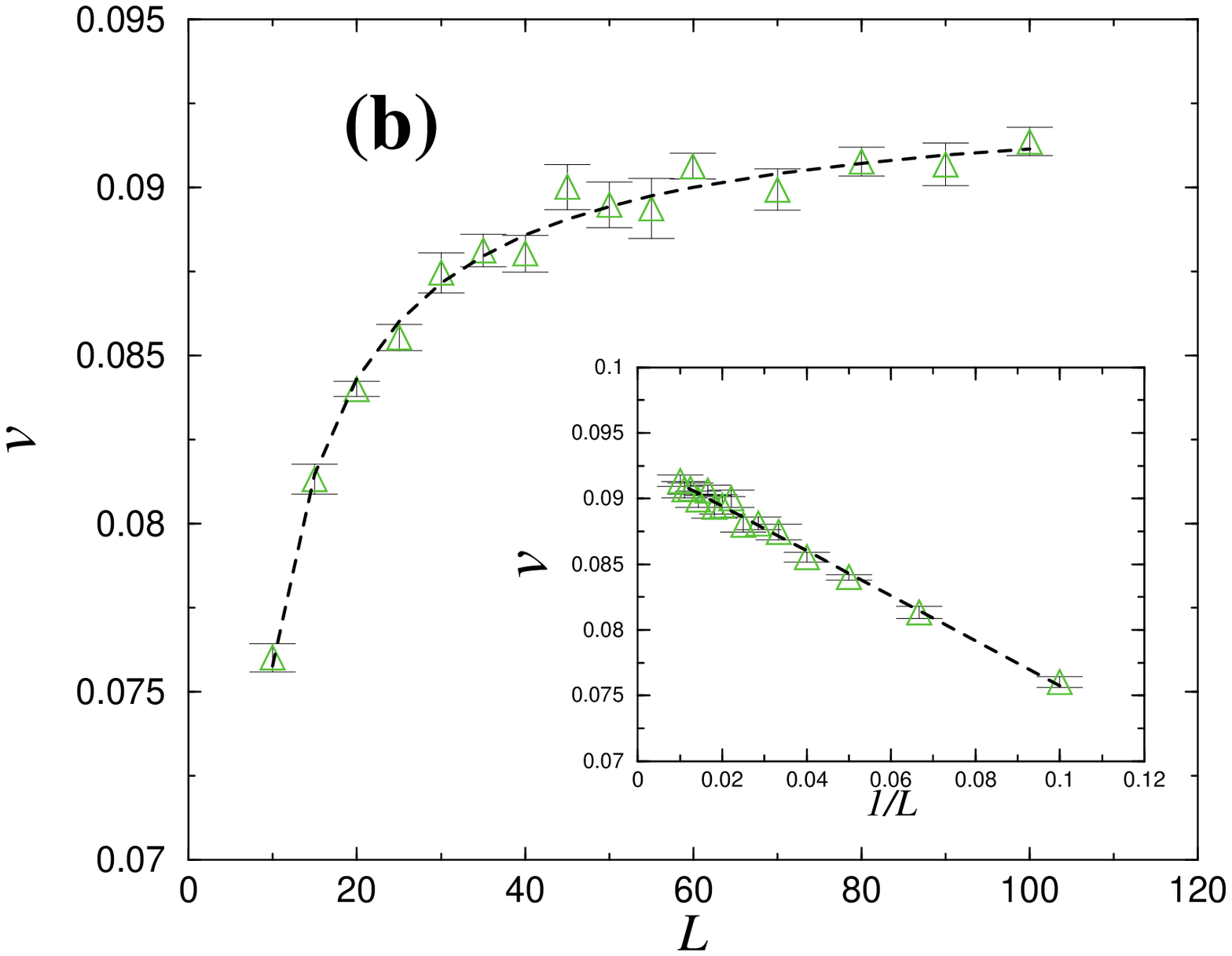}
\vspace*{3.2truecm}
\caption{\small
  (a) Temporal corrections to the asymptotic front velocity
  for $L$$=$$800$ for
  different values of $\alpha_2$, with  $\alpha_1$$=$$0.50$ and $\mu$$=$$0.20$.
  The  horizontal (time) axis is scaled as $t^{-2/3}$, motivated by the form
  of the corrections of the one-dimensional KPZ class [Eq.~(\ref{velocity_corr})]. The inset
  enlarges the region of (a) where the universal temporal corrections
  follow the KPZ behavior. The straight solid lines correspond to
  the linear scaling as a function of $t^{-2/3}$.
  (b) Finite-size corrections to the asymptotic velocity as a function of $L^{-1}$,
  motivated by the universal corrections of the one-dimensional KPZ
  class, for $\alpha_1$$=$$0.50$, $\alpha_2$$=$$0.70$, and $\mu$$=$$0.20$.
  The straight dashed line corresponds to the linear behavior as a function of $1/L$.
  }
\label{fig6}
\end{figure}

The first expression in Eqs.~(\ref{velocity_corr}) indicates that the early
temporal correction to the front's velocity scales as ${\cal
O}(t^{-2/3})$.  Pulled fronts produced in deterministic
reaction-diffusion models exhibit ${\cal O}(t^{-1})$
corrections to the asymptotic velocity \cite{Andow_1990,Holmes_1994}.
A discrete model's pushed front has a larger temporal
correction; effects of time since initiation of invasion (from
$t$$=$$0$ until $t_{\times}$$\sim$$L^z$) persist longer in the
stochastic case. In the simulations we measure $\overline{h}(t)/t$
as it follows the same temporal and system-size scaling as those of
$v(t,L)$ [Eq.~(\ref{velocity_corr})] \cite{OMalley_PRE2006}.
Figure~\ref{fig6}(a) supports the above scaling for our discrete
individual-based two-species invasion front.

More importantly, the second expression in
Eq.~(\ref{velocity_corr}) predicts that after roughening
equilibrates, the invading species' velocity must be corrected
according to the size of the habitat invaded. This correction scales
as ${\cal O}(L^{-1})$. Steady-state velocity increases with
habitat size; longer invading fronts move faster.
Figure~\ref{fig6}(b) and its inset show how
this asymptotic behavior is controlled by system size
$L$. Observed corrections to the velocity
$v^{*}$ decline linearly as a function of $L^{-1}$, in accordance
with the second expression Eq.~(\ref{velocity_corr}). The physical
theory for roughened interfaces explains the effect of habitat size on
frontal velocity, and further gives the form of
the quantitative dependence. These corrections to the
steady-state velocity were observed qualitatively for the basic
contact process \cite{Ellner_1998} and for the Eden model
\cite{Kawasaki_2006}, but neither study addressed either the quantitative
correction for habitat (system) size or the underlying basis for the dynamic behavior:
KPZ roughening of the front.

\subsection{Distribution of extremes in steady-state KPZ fronts: the Airy density}
The distribution of the front's extreme
advance (relative to the mean) provides an additional
strong signature of the underlying universality class of a
stochastic interface. This distribution was only recently obtained
in analytic form \cite{Majumdar_2004,Majumdar_2005}.

As universality arguments imply \cite{Schehr_2006}, the shape of the steady-state
distribution of the front's extreme advance (relative to the mean)
for all (sufficiently large) system sizes and for all models
belonging to the KPZ class will be identical, governed by a single
scale, $\langle \Delta_{max}\rangle \sim L^{1/2}$. For a given
model with a rough front one has
\begin{equation}
P(\Delta_{max},L)= \frac{1}{\langle \Delta_{max}\rangle}\Psi(\Delta_{max}/\langle
\Delta_{max}\rangle)\;,
\label{P_ext_hist2}
\end{equation}
where $\Psi(u)$ is the probability density of the scaled variable
$u\equiv \Delta_{max}/\langle\Delta_{max}\rangle$. For periodic
boundary conditions (employed in our simulations) Majumdar and
Comtet (2004, 2005) obtained the analytic result
\begin{equation}
\Psi(u) = \sqrt{\frac{\pi}{8}}f_A\left(\sqrt{\frac{\pi}{8}}u\right) \;,
\label{P_ext_scaled}
\end{equation}
where $f_A$ is the Airy density function.
A continuous random variable $X$ ($X> $0) has an Airy probability density if:
\begin{equation}
f_A (X) = \frac{2\sqrt{6}}{X^{10/3}} \sum_{k = 1}^\infty e^{-b_k /X^2} b_{k}^{2/3} U(- 5/6, 4/3, b_k/X^2) \;,
\end{equation}
where $b_k = 2{a_k}^3 /27$, and $U(\cdot ,\cdot ,\cdot)$ is the
confluent hypergeometric function.  $a_k$ is the magnitude of the
$k$-th root of the Airy function $Ai(\cdot)$ on the negative real
axis \cite{Abramowitz_1972}.  The Airy distribution arises in a
remarkable number of different applications, including the extremum
of a set of correlated random variables; see Majumdar and Comtet
(2005), Schehr and Majumdar (2006), and Guclu et al. (2007).



\begin{thebibliography}{}

\bibitem[Abramowitz and Stegun 1972]{Abramowitz_1972}
Abramowitz, M., and I. A. Stegun. 1972. Handbook of Mathematical Functions. National Bureau of Standards, Washington, DC.

\bibitem[Allstadt et al. 2007]{Allstadt_2007}
Allstadt, A., T. Caraco, and G. Korniss. 2007. Ecological invasion: spatial clustering and the critical radius.
Evolutionary Ecology Research 9:1--20.

\bibitem[Andow et al. 1990]{Andow_1990}
Andow, D. A., P. M. Kareiva, S. A. Levin, and A. Okubo. 1990. Spread of invading organisms.
Landscape Ecology 4:177--188.

\bibitem[Antal et al. 2001]{Antal_2001}
Antal, T., M. Droz, G. Gy\"orgyi, and Z. R\'acz. 2001.
$1/f$ noise and extreme value statistics.
Physical Review Letters 87:240601, 4p.

\bibitem[Antal et al. 2002]{Antal_2002}
Antal, T., M. Droz, G. Gy\"orgyi, and Z. R\'acz. 2002.
Roughness distribution of $1/f^\alpha$ signals.
Physical Review E 65:046140, 12p.

\bibitem[Antonovics et al. 2006]{Antonovics_2006}
Antonovics, J., A. J. McKane, and T. J. Newman. 2006.
Spatiotemporal dynamics in marginal populations.
American Naturalist 167:16--27.

\bibitem[Aronson and Weinberger 1978]{Aronson_1978}
Aronson, D. G., and H. F. Weinberger. 1978. Multidimensional nonlinear diffusion arising in population genetics.
Advances in Mathematics 30:33--76.

\bibitem[Aylor 2003]{Aylor_2003}
Aylor, D.E. 2003. Spread of plant disease on a continental scale:
role of aerial dispersal of pathogens. Ecology 84:1989--1997.

\bibitem[Barab\'asi and Stanley 1995]{Barabasi_1995}
Barab\'{a}si, A.-L., and H. E. Stanley. 1995. Fractal Concepts in Surface Growth. Cambridge University Press,
Cambridge.

\bibitem[Berman 1964]{Berman_1964}
Berman, S. M. 1964.
Limit theorems for the maximum term in stationary sequences.
Annals of Mathematical Statistics 35:502--516.

\bibitem[ben Avraham 1998]{bAvraham_1998}
ben-Avraham, D. 1998. Fisher waves in the diffusion limited coalescence process. Physics Letters A 247:53--58.

\bibitem[Bjornstad et al. 2002]{Bjornstad_2002}
Bjornstad, O.N., M. Peltonin, A.M. Liebhold, and W. Baltensweiler. 2002.
Waves of larch budmoth outbreaks in the European Alps. Science 298:1020--1023.

\bibitem[Blythe and Evans 2001]{Blythe_2001}
Blythe, R. A., and M. R. Evans. 2001.
Slow crossover to Kardar-Parisi-Zhang scaling.
Physical Review E 64:051101, 5p.

\bibitem[Br\'{u} et al. 2003]{Bru_2003}
Br\'u, A., S. Albertos, J. L. Subiza, J. L. Garc\'ia-Asenjo, and I. Br\'u. 2003.
The universal dynamics of tumor growth.
Biophysics Journal 85:2948--2961.

\bibitem[Cain et al. 1995]{Cain_1995}
Cain, M.L., S.W. Pacala, J.A. Silander Jr., and M.-J. Fortin. 1995.
Neighborhood models of clonal growth in the white clover \emph{Trifolium repens}.
American Naturalist 145:888--917.

\bibitem[Cannas et al. 2006]{Cannas_2006}
Cannas, S.A., D.E. Marco, and M.A. Montemurro. 2006. Long range dispersal and spatial pattern formation
in biological invasions. Mathematical Biosciences 203:155--170.

\bibitem[Cantrell and Cosner 1991]{Cantrell_1991}
Cantrell, R. S., and C. Cosner. 1991. The effect of spatial heterogeneity in population dynamics.
Journal of Mathematical Biology 29:315--338.

\bibitem[Caraco et al. 2002]{Caraco_2002}
Caraco, T., S. Glavanakov, G. Chen, J. E. Flaherty, T. K. Ohsumi, and B. K. Szymanski. 2002. Stage-structured infection transmission and a spatial epidemic: a model for Lyme disease. American Naturalist 160:348--359.

\bibitem[Cardy 1996]{Cardy_1996}
Cardy, J. 1996. Scaling and Renormalization in Statistical Physics. Cambridge University Press,
Cambridge.

\bibitem[Clark et al. 1998]{Clark_1998}
Clark, J.S., C. Fastie, G. Hurtt, S.T. Jackson, C. Johnson, G.A. King, M. Lewis, J. Lynch,
S. Pacala, C. Prentice, E.W. Schupp. T. Webb III, and P. Wyckoff. 1998. Reid's paradox of rapid plant migration.
BioScience 48:13--24.

\bibitem[Clark et al. 2001]{Clark_2001}
Clark, J. S., M. Lewis, and L. Horvath. 2001. Invasion by extremes: population spread with variation in dispersal and reproduction. American Naturalist 157:537--554.

\bibitem[Clark et al. 2003]{Clark_2003}
Clark, J. S., M. Lewis, J. S. McLachlan, and J. HilleRisLambers. 2003. Estimating population spread: what can we forecast and how well? Ecology 84:1979--1988.

\bibitem[Comins and Noble 1985]{Comins_1985}
Comins, H. N., and I. R. Noble. 1985. Dispersal, variability, and transient niches: species coexistence in a uniformly variable environment. American Naturalist 126:706--723.

\bibitem[Connolly and Muko 2003]{Connonlly_2003}
Connolly, S.R., and S. Muko. 2003. Space preemption, size-dependent competition and
the coexistence of clonal growth forms. Ecology 84:2979--2988.

\bibitem[DAntonio 1993]{DAntonio_1993}
D'Antonio, C. M. 1993. Mechanisms controlling invasion of coastal plant communities by the alien succulent \emph{Carpobrotus} \emph{edulis}. Ecology 74:83--95.

\bibitem[DeAngelis and Gross 1992]{DeAngelis_1992}
DeAngelis, D. L., and L. J. Gross (Eds.) 1992. Individual-Based Models
and Approaches in Ecology. Routledge, Chapman and Hall, New York.

\bibitem[Doering et al. 2003]{Doering_2003}
Doering C. R., C. Mueller C, and P. Smereka. 2003.
Interacting particles, the stochastic Fisher-Kolmogorov-Petrovsky-Piscounov equation, and duality.
Physica A 325:243--259.

\bibitem[Doi 1976]{Doi_1976}
Doi, M. 1976. Stochastic theory of diffusion-controlled reaction.
Journal of Physics A 9:1479--1495.

\bibitem[Durrett and Levin 1994a]{Durrett_1994ptrsl}
Durrett, R., and S. A. Levin. 1994a. Stochastic spatial models: a user's guide to ecological applications.
Philosophical Transactions of the Royal Society, London B 343:329--350.

\bibitem[Durrett and Levin 1994b]{Durrett_1994tpb}
Durrett, R., and S. A. Levin. 1994b. The importance of being discrete (and spatial).
Theoretical Population Biology 46:363--394.

\bibitem[Dwyer 1992]{Dwyer_1992}
Dwyer, G. 1992. On the spatial spread of insect pathogens:
theory and experiment. Ecology 73:479--494.

\bibitem[Dwyer and Elkinton 1995]{Dwyer_1995}
Dwyer, G., and S. Elkinton. 1995. Host dispersal and the spatial spread of insect pathogens. Ecology 76:1262--1275.

\bibitem[Dwyer and Morris 2006]{Dwyer_2006}
Dwyer, G., and W. F. Morris. 2006. Resource-dependent dispersal and the speed of biological invasions.
American Naturalist 167:165--176.

\bibitem[Eden 1961]{Eden_1961}
Eden, M. 1961. A two-dimensional growth process. In J. Neyman (Ed.)
4th Berkeley Symposium on Mathematical Statistics and Probability,
vol 4, pp. 223--239. University of California Press, Berkeley, CA.

\bibitem[Ellner et al. 1998]{Ellner_1998}
Ellner, S. P., A. Sasaki, Y. Haraguchi, and H. Matsuda. 1998.
Speed of invasion in lattice population models: pair-edge approximation.
Journal of Mathematical Biology 36:469--484.

\bibitem[Elton 1958]{Elton_1958}
Elton, C. S. 1958. The Ecology of Invasions by Animals and Plants. Methuen Press, London.

\bibitem[Escudero et al. 2004]{Escudero_2004}
Escudero, C., J. Buceta, F. J de la Rubia, and K. Lindenberg. 2004. Extinction in population dynamics.
Physical Review E 69:021908, 9 p.

\bibitem[Family and Vicsek 1985]{Family_1985}
Family, F., and T. Vicsek. 1985.
Scaling of the active zone in the Eden process on percolation networks and the ballistic deposition model.
Journal of Physics A 18:L75--L81.

\bibitem[Ferrandino 1996]{Ferrandino_1996}
Ferrandino, F.J. 1996. Length scale of disease spread: fact or artifact of experimental geometry?
Phytopathology 86:806--811.

\bibitem[Ferreira and Alves 2006]{Ferreira_2006}
Ferreira Jr., S. C., and S. G. Alves. 2006. Pitfalls in the determination of the universality class of radial clusters. Journal of Statistical Mechanics 11:P11007, 11 p.

\bibitem[Fisher et al. 2001]{Fisher_2001}
Fisher, M.C., G.L. Koenig, T.J. White, G. Sans-Blas, R. Negroni, I.G. Alvarez, B. Wanke, and
J.W. Taylor. 2001. Biogeographic range expansion into South America by
\emph{Coccidioides immitis} mirrors New World patterns of human migration.
Proceedings of the National Academy of Science USA 98:4558--4562.

\bibitem[Fisher 1937]{Fisher_1937}
Fisher, R. A. 1937. The wave of advance of advantageous genes. Annals of Eugenics London 7:355--369.

\bibitem[Fisher 1928]{Fisher_1928}
Fisher, R. A., and L. H. C. Tippett. 1928. The frequency distribution of the largest or smallest member of a sample. Proceedings of the Cambridge Philosophical Society 24:180--191.

\bibitem[Foltin et al. 1994]{Foltin_1994}
Foltin, G., K. Oerding, Z. R\'{a}cz, R. L. Workman, and R. K. P. Zia. 1994.
Width distribution for random-walk interfaces.
Physical Review E 50:R639--R642.

\bibitem[Frantzen and van den Bosch 2000]{Frantzen_2000}
Frantzen, J., and F. van den Bosch. 2000. Spread of organisms: can travelling
and dispersive waves be distinguished? Basic and Applied Ecology 1:83--91.

\bibitem[Galambos 1987]{Galambos_1987}
Galambos, J. 1987. The Asymptotic Theory of Extreme Order Statistics (2nd ed).
Krieger Publishing Co., Malabar, FL.

\bibitem[Galambos et al. 1994]{Galambos_1994}
Galambos, J., J. Lechner, and E. Simin (Eds.). 1994. Extreme Value Theory and Applications. Kluwer, Dordrecht.

\bibitem[Gandhi et al. 1999]{Gandhi_1999}
Gandhi, A., S. Levin, and S. Orszag. 1999. Nucleation and relaxation from meta-stability in spatial ecological models. Journal of Theoretical Biology 200:121--146.

\bibitem[Gardiner 1985]{Gardiner_1985}
Gardiner, C. W. 1985. Handbook of Stochastic Methods for Physics,
Chemistry and the Natural Sciences, 2nd ed., Springer, Berlin.

\bibitem[Guclu and Korniss 2004]{Guclu_2004}
Guclu, H., and G. Korniss. 2004.
Extreme fluctuations in small-worlds with relaxational dynamics.
Physical Review E 69:065104(R), 4p.

\bibitem[Guclu et al. 2007]{Guclu_2007}
Guclu, H., G. Korniss, and Z. Toroczkai. 2007.
Extreme fluctuations in noisy task-completion landscapes on scale-free networks.
Chaos 17:026104, 13p.

\bibitem[Gumbel 1928]{Gumbel_1958}
Gumbel, E. J. 1958. Statistics of Extremes. Columbia Univerisity Press, New York.

\bibitem[Halpin-Healy and Zhang 1995]{Halpin_1995}
Halpin-Healy, T., and Y.-C. Zhang. 1995. Kinetic roughening phenomena, stochastic growth, directed polymers and all that. Aspects of multidisciplinary statistical mechanics. Physics Reports 254:215--414.

\bibitem[Harris 1974]{Harris_1974}
Harris, T. E. 1974. Contact interaction on a lattice. Annals of Probability 2:969--988.

\bibitem[Hastings et al. 2005]{Hastings_2005}
Hastings, A., K. Cuddington, K.F. Davies, C.J. Dugaw, S. Elmendorf, A. Freestone,
S. Harrison, M. Holland, J. Lambrinos, U. Malvadkar, B.A. Melbourne, K. Moore,
C. Taylor, and D. Thomson. 2005. The spatial spread of invasions: new developments in
theory and evidence. Ecology Letters 8:91--101.

\bibitem[Hinrichsen 2000]{Hinrichsen_2000}
Hinrichsen, H. 2000. Non-equilibrium critical phenomena and phase transitions into absorbing states.
Advances in Physics 49:815--958.

\bibitem[Holmes et al. 1994]{Holmes_1994}
Holmes, E.E., M. A. Lewis, J. E. Banks, and R. R. Veit. 1994. Partial differential equations in ecology: spatial interactions and population dynamics. Ecology 75:17--29.

\bibitem[Holway 1998]{Holway_1998}
Holway, D. A. 1998. Factors governing rate of invasion: a natural experiment using Argentine ants. Oecologia
115:206--212.

\bibitem[Hoopes and Hall 2002]{Hoopes_2002}
Hoopes, M.F., and L. M. Hall. 2002. Edaphic factors and competition affect pattern formation and invasion in a California grassland. Ecological Applications 12:24--39.

\bibitem[Hosono 1998]{Hosono_1998}
Hosono, Y. 1998. The minimal speed of travelling fronts for a diffusive Lotka-Volterra competition model.
Bulletin of Mathematical Biology 60:435--448.

\bibitem[Jullien and Botet 1985a]{Jullien_PRL1985}
Jullien, R., and R. Botet. 1985a. Surface thickness in the Eden model. Physical
Review Letters 54:2055.

\bibitem[Jullien and Botet 1985b]{Jullien_JPA1985}
Jullien, R., and R. Botet. 1985b. Scaling properties of the surface of the Eden model.
Journal of Physics A 18:2279--2287.

\bibitem[Kardar et al. 1986]{Kardar_1986}
Kardar, M., G. Parisi, and Y.-C. Zhang. 1986. Dynamic scaling of growing interfaces.
Physical Review Letters 56:889--892.

\bibitem[Kawasaki et al. 2006]{Kawasaki_2006}
Kawasaki, K., F. Takasu, H. Caswell, and N. Shigesada. 2006. How does stochasticity in colonization accelerate the speed of invasion in a cellular automaton model? Ecological Research 21:334--345.

\bibitem[Kert\'{e}sz and Wolf 1988]{Kertesz_1988}
Kert\'{e}sz, J., and D. E. Wolf. 1988. Noise reduction in Eden models: II. Surface structure and intrinsic width.
Journal of Physics A: Math General 21:747--761.

\bibitem[Kolmogorov et al. 1937]{Kolmogorov_1937}
Kolmogorov, A., N. Petrovsky, and N. S. Pishkounov. 1937. A study of the equation of diffusion with increase in the quantity of matter, and its application to a biological problem. Moscow University
Bulletin of Mathematics 1:1--25.

\bibitem[Korniss 1997]{Korniss_1997}
Korniss, G. 1997. Structure factors and their distributions in driven two-species models.
Physical Review E 56:4072--4084.

\bibitem[Korniss and Caraco 2005]{Korniss_JTB2005}
Korniss, G., and T. Caraco. 2005. Spatial dynamics of invasion: the geometry of introduced species.
Journal of Theoretical Biology 233:137--150.

\bibitem[Korniss et al. 2003]{Korniss_2003}
Korniss, G., M. A. Novotny, H. Guclu, Z. Toroczkai, and P. A. Rikvold (2003)
Suppressing roughness of virtual times in parallel discrete-event simulations.
Science 299:677--679.

\bibitem[Korniss et al. 2000]{Korniss_2000}
Korniss, G., Z. Toroczkai, M. A. Novotny, and P. A. Rikvold (2000)
From massively parallel algorithms and fluctuating time horizons to nonequilibrium surface growth.
Physical Review Letters 84:1351--1354.

\bibitem[Kot et al. 1996]{Kot_1996}
Kot, M., M. A. Lewis, and P. van den Driessche. 1996. Dispersal data and the spread of invading organisms.
Ecology 77:2027--2042.

\bibitem[Krug and Meakin 1990]{Krug_1990}
Krug, J., and P. Meakin. 1990. Universal finite-size effects in the rate of growth processes.
Journal of Physics A 23:L987--L994.

\bibitem[Lewis 1997]{Lewis_1997}
Lewis, M. A. 1997. Variability, patchiness, and jump dispersal in the spread of an
invading population. In D. Tilman, P. Kareiva (Eds.) Spatial Ecology: The Role of Space in
Population Dynamics and Interspecific Interactions, pp. 46--69. Princeton University Press,
Princeton.

\bibitem[Lewis 2000]{Lewis_2000}
Lewis, M. A. 2000. Spread rate for a nonlinear stochastic invasion.
Journal of Mathematical Biology 41:430--454.

\bibitem[Lewis et al. 2002]{Lewis_2002}
Lewis, M. A., B. Li, and H. F. Weinberger. 2002. Spreading speed and linear determinacy for
two-species competition models. Journal of Mathematical Biology 45:219--233.

\bibitem[Lockwood et al. 2007]{Lockwood_2007}
Lockwood, J. L., M. F. Hoopes, and M. Marchetti. 2007. Invasion Ecology. Blackwell Publishing, Malden, MA.

\bibitem[Majumdar and Comtet 2004]{Majumdar_2004}
Majumdar, S. N., and A. Comtet. 2004. Exact maximal height distribution of fluctuation interfaces.
Physical Review Letters 92:225501, 4 p.

\bibitem[Majumdar and Comtet 2005]{Majumdar_2005}
Majumdar, S. N., and A. Comtet. 2005.
Airy distribution function: from the area under a Brownian excursion to the maximal height of fluctuating interfaces.
Journal of Statistical Physics 119:776--826.

\bibitem[McKane and Newman 2004]{McKane_2004}
McKane, A. J., and T. J. Newman. 2004. Stochastic models in population biology and their deterministic analogues.
Physical Review E 70:041902, 19 p.

\bibitem[Minogue and Fry 1983]{Minogue_1983}
Minogue, K. P., and W. E. Fry. 1983.
Models for the spread of plant disease: some experimental results.
Phytopathology 73: 1173--1176.

\bibitem[Mollison and Levin 1995]{Mollison_1995}
Mollison, D., and S. A. Levin. 1995.
Spatial dynamics of parasitism. In B. T. Grenfell and  A. P. Dobson (Eds.)
Ecology of Infectious Diseases in Natural Populations, pp. 384--398.
Cambridge University Press, Cambridge.

\bibitem[Moro 2001]{Moro_2001}
Moro, E. 2001. Internal fluctuations effects on Fisher waves. Physical Review Letters 87:238303, 4 p.

\bibitem[Moro 2003]{Moro_2003}
Moro, E. 2003. Emergence of pulled fronts in fermionic microscopic particle models.
Physical Review E 68:025102, 4 p.

\bibitem[Murray 2003]{Murray_2003}
Murray, J. D. 2003. Mathematical Biology, vol 2. Springer, New York.

\bibitem[Nash et al. 1995]{Nash_1995}
Nash, D.R., D.J.L. Agassiz, H.C.J. Godfray, and J.H. Lawton. 1995.
The pattern of spread of invading species: two leaf-mining moths
colonizing Great Britain. Journal of Animal Ecology 64:225--233.

\bibitem[Neubert and Caswell 2000]{Neubert_2000}
Neubert, M. G., and H. Caswell. 2000.
Demography and dispersal: calculation and sensitivity analysis of invasion speed for structured populations.
Ecology 81:1613--1628.

\bibitem[Oborny et al. 2005]{Oborny_2005}
Oborny, B., G. Mesz\'{e}na, and G. Szab\'{o}. 2005. Dynamics of populations on the verge of extinction.
Oikos 109:291--296.

\bibitem[O'Malley et al. 2005]{OMalley_SPIE2005}
O'Malley, L., A. Allstadt, G. Korniss, and T. Caraco. 2005.
Nucleation and global time scales in ecological invasion under preemptive competition.
In N. G. Stocks, D. Abbott, and R. P. Morse (Eds.) Fluctuations and Noise in Biological,
Biophysical, and Biomedical Systems III, pp. 117--124.  SPIE, Pullman, WA.

\bibitem[O'Malley et al. 2006a]{OMalley_TPB2006}
O'Malley, L., J. Basham, J. A. Yasi, G. Korniss, A. Allstadt, and T. Caraco. 2006a.
Invasive advance of an advantageous mutation: nucleation theory.
Theoretical Population Biology 70:464--478.

\bibitem[O'Malley et al. 2006b]{OMalley_PRE2006}
O'Malley, L., B. Kozma, G. Korniss, Z. R\'{a}cz, and T. Caraco. 2006b.
Fisher waves and front propagation in a two-species invasion model with preemptive competition.
Physical Review E 74:041116, 7 p.

\bibitem[O'Malley et al. 2009]{OMalley_SPRINGER}
O'Malley, L., B. Kozma, G. Korniss, Z. R\'{a}cz, and T. Caraco. 2009.
Fisher waves and the velocity of front propagation in a two-species invasion model with preemptive competition.
In Computer Simulation Studies in Condensed Matter Physics XIX,
D. P. Landau, S. P. Lewis, and H.-B. Sch\"{u}ttler (Eds.),
Springer Proceedings in Physics Vol. 123, pp. 73--78. Springer, Heidelberg.

\bibitem[Parker and Reichard 1998]{Parker_1998}
Parker, I. M., and S. H. Reichard. 1998.  Critical issues in invasion biology for conservation science.
In P. L. Fieldler and P. M. Kareiva (Eds.) Conservation Biology, 2nd ed, pp. 283--305.
Chapman and Hall, New York.

\bibitem[Pechenik and Levine 1999]{Pechenik_1999}
Pechenik, L., and H. Levine. 1999. Interfacial velocity corrections due to multiplicative noise.
Physical Review E 59:3893--3900.

\bibitem[Peliti 1985]{Peliti_1985}
Peliti, L. 1985. Path integral approach to birth-death processes on a lattice.
Journal of Physics (Paris) 46:1469--1483.

\bibitem[Pimentel et al. 2000]{Pimentel_2000}
Pimentel, D., L. Lach, R. Zuniga, and D. Morrison. 2000.
Environmental and economic costs of nonindigenous species in the United States.
Bioscience 50:53--65.

\bibitem[Plischke and R\'acz 1985]{Plischke_1985}
Plischke, M., and Z. R\'acz. 1985. Dynamic scaling and the surface structure of Eden clusters.
Physical Review A 32:3825--3828.

\bibitem[Plischke et al. 1987]{Plischke_1987}
Plischke, M., Z. R\'acz, and D. Liu (1987)
Time-reversal invariance and universality of two-dimensional growth models.
Physical Review B 35:3485--3495.

\bibitem[R\'{a}cz and G\'{a}lfi 1988]{Racz_1988}
R\'{a}cz, Z., and L. G\'{a}lfi. 1988.
Properties of the reaction front in an \emph{A} + \emph{B} $\rightarrow$ \emph{C} type reaction-diffusion process.
Physical Review A 38:3151--3154.

\bibitem[Raychaudhuri et al. 2001]{Raychaudhuri_2001}
Raychaudhuri, S., M. Cranston, C. Przybyla, and Y. Shapir. 2001.
Maximal height scaling of kinetically growing surfaces.
Physical Review Letters 87:136101, 4 p.

\bibitem[Rosenzweig 2001]{Rosenzweig_2001}
Rosenzweig, M. L. 2001. The four questions: what does the introduction of exotic species do to diversity?
Evolutonary Ecology Research 3:361--371.

\bibitem[Ruesink et al. 1995]{Ruesink_1995}
Ruesink, J. L., I. M. Parker, M. J. Groom, and P. M. Kareiva. 1995.
Reducing the risks of nonindigenous introductions:
guilty until proven innoent. BioScience 45:465-477.

\bibitem[Ruiz et al. 2000]{Ruiz_2000}
Ruiz, G.M., T.K. Rawlings, F.C Dobbs, A. Huq, and R. Colwell. 2000. Global spread of microorganisms by ships.
Nature 408:49.

\bibitem[Schehr and Majumdar 2006]{Schehr_2006}
Schehr, G., and S. N. Majumdar. 2006.
Universal asymptotic statistics of a maximal relative height in one-dimensional solid-on-solid models.
Physical Review E 73:056103, 10 p.

\bibitem[Schmittmann and Zia 1995]{Schmittmann_1995}
Schmittman, B., and R. K. P. Zia. 1995. Statistical Mechanics of Driven Diffusive Systems.
Phase Transitions and Critical Phenomena, vol. 17. Academic Press, New York.

\bibitem[Schwinning and Parsons 1996]{Schwinning_1996}
Schwinning, S., and A. J. Parsons. 1996. A spatially explicit population model of
stoloniferous N-fixing legumes in mixed pasture with grass.
Journal of Ecology 84:815--826.

\bibitem[Shigesada and Kawasaski 1997]{Shigesada_97}
Shigesada, N., and K. Kawasaki. 1997. Biological Invasions: Theory and Practice.
Oxford University Press, Oxford, UK.

\bibitem[Shigesada et al. 1995]{Shigesada_1995}
Shigesada, N., K. Kawasaki, and Y. Takeda. 1995. Modeling stratified diffusion in biological invasions.
American Naturalist 146:229--251.

\bibitem[Simberloff et al. 2002]{Simberloff_2002}
Simberloff, D., M. A. Relva, and M. Nu\~{n}ez. 2002.
Gringos en el bosque: introduced tree invasion in a native \emph{Nothofagus}/\emph{Austrocedrus} forest.
Biological Invasions 4:35--53.

\bibitem[Silvertown et al. 1994]{Silvertown_1994}
Silvertown, J., C. E. M. Lines, and M. P. Dale. 1994. Spatial competition between grasses -- rates
of mutual invasion between four species and the interaction with grazing.
Journal of Ecology 82:31--38.

\bibitem[Snyder 2003]{Snyder_2003}
Snyder, R. E. 2003. How demographic stochasticity can slow biological invasions. Ecology 84:1333--1339.

\bibitem[Tainaka et al. 2004]{Tainaka_2004}
Tainaka, K., M. Kushida, Y. Itoh, and J. Yoshimura. 2004.
Interspecific segregation in a lattice ecosystem with intraspecific competition.
Journal Physics Society of Japan 73:2914--2915.

\bibitem[Thomson and Ellner 2003]{Thomson_2003}
Thomson, N. A., and S. P. Ellner. 2003. Pair-edge approximation for heterogeneous lattice population models.
Theoretical Population Biology 64:270--280.

\bibitem[van Baalen and Rand 1998]{vBaalen_1998}
van Baalen, M., and D. A. Rand. 1998. The unit of selection in viscous populations and the evolution of altruism.
Journal of Theoretical Biology 193:631--648.

\bibitem[van den Bosch et al. 1992]{vdBosch_1992}
van den Bosch, F., R. Hengeveld, and J. A. J. Metz. 1992. Analysing the velocity of animal range expansion.
Journal of Biogeography 19:135--150.

\bibitem[van Saarloos 2003]{vSaarloos_2003}
van Saarloos, W. 2003. Front propagation into unstable states. Physics Reports 386:29--222.

\bibitem[van Kampen 1974]{vKampen_1976}
van Kampen, N. G. 1976. The expansion of the master equation.
Advances in Chemical Physics 34:245--309.

\bibitem[van Kampen 1981]{vKampen_1981}
van Kampen, N. G. 1981.
Stochastic Processes in Physics and Chemistry.
Elsevier, Amsterdam.

\bibitem[Weinberger et al. 2002]{Weinberger_2002}
Weinberger, H. F., M. A. Lewis, and B. T. Li. 2002. Analysis of linear determinacy for spread in cooperative models.
Journal of Mathematical Biology 45:183--218.

\bibitem[Wilson 1998]{Wilson_1998}
Wilson, W. 1998. Resolving discrepancies between deterministic population models and individual-based simulations.
American Naturalist 151:116--134.

\bibitem[Wilson et al. 1993]{Wilson_1993}
Wilson, W., A. M. de Roos, and E. McCauley. 1993.
Spatial instabilities within the diffusive Lotka-Volterra system:
individual-based simulation results. Theoretical Population Biology 43:91--127.

\bibitem[Yasi et al. 2006]{Yasi_2006}
Yasi, J., G. Korniss, and T. Caraco. 2006.
Invasive allele spread under preemptive competition.
In Computer Simulation Studies in Condensed Matter Physics XVIII,
D. P. Landau, S. P. Lewis, and H.-B. Sch\"{u}ttler (Eds.),
Springer Proceedings in Physics Vol. 105, pp. 165--169. Springer, Heidelberg.


\bibitem[Yurkonis and Meiners 2004]{Yurkonis_2004}
Yurkonis, K.A., and Meiners, S.J. 2004. Invasion impacts local species turnover
in a successional system. Ecology Letters 4:764--769.


\end{thebibliography}
\end{document}